\documentclass[final]{siamltex}
\usepackage{epsfig}
\usepackage{graphicx}
\usepackage{amsmath}
\usepackage{amssymb}
\usepackage{setspace}
\usepackage{graphicx}
\graphicspath{{pics/}{figs/}}
\usepackage{graphicx}

\newtheorem{defn}{\noindent $\mathbf{Definition}$}[section]

\newtheorem{thm}[defn]{$\mathbf{Theorem}$}

\title{Teichm\"uller Extremal Mapping and its Applications to Landmark Matching Registration}
\author{Lok Ming Lui, Ka Chun Lam, Shing-Tung Yau and Xianfeng Gu}

\begin{document}

\maketitle

\begin{abstract}
Registration, which aims to find an optimal 1-1 correspondence between shapes, is an important process in different research areas. Conformal mappings have been widely used to obtain a diffeomorphism between shapes that minimizes angular distortion. Conformal registrations are beneficial since it preserves the local geometry well. However, when landmark constraints are enforced, conformal mappings generally do not exist. This motivates us to look for a unique landmark matching quasi-conformal registration, which minimizes the conformality distortion. Under suitable condition on the landmark constraints, a unique diffeomporphism, called the {\it Teichm\"uller extremal mapping} between two surfaces can be obtained, which minimizes the maximal conformality distortion. In this paper, we propose an efficient iterative algorithm, called the {\it Quasi-conformal (QC) iterations}, to compute the Teichm\"uller mapping. The basic idea is to represent the set of diffeomorphisms using Beltrami coefficients (BCs), and look for an optimal BC associated to the desired Teichm\"uller mapping. The associated diffeomorphism can be efficiently reconstructed from the optimal BC using the Linear Beltrami Solver(LBS). Using BCs to represent diffeomorphisms guarantees the diffeomorphic property of the registration. Using our proposed method, the Teichm\"uller mapping can be accurately and efficiently computed within 10 seconds. The obtained registration is guaranteed to be bijective. The proposed algorithm can also be extended to compute Teichm\"uller mapping with soft landmark constraints. We applied the proposed algorithm to real applications, such as brain landmark matching registration, constrained texture mapping and human face registration. Experimental results shows that our method is both effective and efficient in computing a non-overlap landmark matching registration with least amount of conformality distortion.
\end{abstract}



\pagestyle{myheadings}
\thispagestyle{plain}
\markboth{Lok Ming Lui et al.}{Teichm\"uller mapping and applications}

\section{Introduction}
Registration refers to the process of finding an optimal one-to-one correspondence between images or surfaces. It has been extensively applied to different areas such as medical imaging, computer graphics and computer visions. For example, in medical imaging, registration is always needed for statistical shape analysis, morphometry and processing of signals on brain surfaces (e.g., denoising or filtering). While in computer graphics, surface registration is needed for texture mapping, which align each vertex to a position of the texture image, to improve the visualization of the surface mesh. Developing an effective algorithm for registration is therefore very important.

Conformal mappings have been widely used to obtain smooth 1-1 correspondences between different domains that minimize angular distortions. Conformal mappings are advantageous since it preserves the local geometry well. According to the conformal geometry, given two simply-connected domains, there always exists a unique conformal mapping between them up to a Mobi\"us transformation. However, the existence of conformal mappings cannot be guaranteed in general situations. For example, conformal mapping between two multiply-connected domains usually does not exist. Besides, in practical situation, obtaining a registration that matches landmark features consistently are often required. When landmark constraints are enforced, the existence of conformal mappings cannot be guaranteed. This motivates us to look for a unique landmark-matching registration, which minimizes the conformality distortion as much as possible.

Under suitable conditions on the landmark constraints, there exists a unique quasi-conformal mapping, called the {\it Teichm\"uller extremal mapping}, which minimizes the maximal conformality distortion. To compute this Teichm\"uller mapping, we propose in this paper an efficient and effective iterative algorithm, which is called the {\it Quasi-conformal (QC) iterations}. The basic idea is to represent the set of diffeomorphisms using Beltrami coefficients (BCs), and look for an optimal BC associated to the desired Teichm\"uller mapping. The associated diffeomorphism can be efficiently reconstructed from the optimal BC using the Linear Beltrami Solver(LBS). Given a set of landmark constraints, the algorithm is able to determine an optimal 1-1 correspondence (including the boundary correspondence in the case of open surfaces) between shapes automatically, which minimizes the conformality distortion. Besides, the proposed algorithm can also compute extremal mapping with soft landmark constraints. It becomes necessary when landmark features cannot be accurately located, and hence it is better to compute registration with landmarks approximately (but not exactly) matched.  Another major advantage of using Teichm\"uller mappings for landmark matching registrations is that the bijectity (1-1, onto) of the registrations can be guaranteed. Obtaining a bijective landmark matching registration is generally difficult, especially when a large number of landmark constraints are enforced.  Using our proposed method, a bijective Teichm\"uller mapping can be accurately and efficiently computed within 10 seconds. To test the effectiveness of our method, we applied the proposed algorithm to real applications, such as brain registration, constrained texture mapping and human face registration. Experimental results shows that our method is both effective and efficient in computing a non-overlap landmark matching registration with least conformality distortion.

In short, the contributions of this paper are three-folded. Firstly, we propose an efficient algorithm for obtaining the unique {\it Teichm\"uller extremal mapping} between shapes with landmark constraints enforced. The mapping is guaranteed to be bijective and minimizes the maximal conformality distortion. Secondly, we propose an algorithm to compute Teichm\"uller extremal mapping with soft landmark constraints. Landmarks are not exactly matched, but less conformality distortion will be introduced. Thirdly, we apply the proposed algorithms to real applications, namely, constrained texture mapping, medical image registration and human face registration.

\section{Previous work}
In this section, we will introduce some previous works closely related to our paper.

Surface parameterization and registration have been extensively studied, for which different kinds of bijective surface maps have been proposed. Conformal registration, which minimizes angular distortion, have been widely used to obtain a smooth 1-1 correspondence between surfaces \cite{Fischl2, Gu1, Gu3, Haker, Hurdal, Gu2}. For example, Hurdal et al. \cite{Hurdal} proposed to compute the conformal parameterizations using circle packing and applied it to registration of human brains. Gu et al. \cite{Gu1, Gu3, Gu2} proposed to compute the conformal parameterizations of Riemann surfaces for registration using harmonic energy minimization and holomorphic 1-forms. Later, the authors proposed the curvature flow method to compute conformal parameterizations of high-genus surfaces onto their universal covering spaces \cite{JinRicci,YLYangRicci,LuiVSRicci}. Conformal registration is advantageous for it preserves the local geometry well.

Sometimes, deformations between objects might not not conformal. Instead, certain amount of angular distortion could be introduced. To tackle with this situation, quasi-conformal mappings are proposed to obtain smooth 1-1 correspondence with bounded conformality distortion. Lui et al. \cite{LuiBHFHP} proposes to compute quasi-conformal registration between hippocampal surfaces which matches geometric quantities (such as curvatures) as much as possible. A method called the Beltrami Holomorphic flow is used to obtain the optimal Beltrami coefficient associated to the registration \cite{LuiBHF}. Beltrami coefficient has been applied to represent general surface homeomorphisms, which is comparatively easier to manipulate than 3D coordinate functions. Using Beltrami representation, compression of surface maps has been proposed \cite{LuiCompression}, which can be applied for video compression \cite{LuiBeltramirepresentation}. Wei et al. \cite{Weiface} also proposes to compute quasi-conformal mapping for feature matching face registration. The Beltrami coefficient associated to a landmark points matching parameterization is approximated. However, either exact landmark matching or the bijectivity of the mapping cannot be guaranteed, especially when very large deformations occur. In order to compute quasi-conformal mapping from the Beltrami coefficients effectively. Quasi-Yamabe method is introduced, which applies the curvature flow method to compute the quasi-conformal mapping \cite{LuiQuasiYamabe}. The algorithm can deal with surfaces with general topologies. Later, quasi-conformal mapping, which minimizes the conformality distortion, has been studied. Zorin et al. \cite{Zorin} proposes an algorithm to compute the extremal quasi-conformal mapping between connected domains with given Dirichlet condition defined on the whole boundaries. The extremal mapping is obtained by minimizing a least square Beltrami energy, which is non-convex. The algorithm can obtain an extremal mapping when initialization is carefully chosen. However, the convergence to the global minimum cannot be guaranteed.

Most of the above registration algorithms using conformal and quasi-conformal mappings cannot match feature landmarks, such as sulcal landmarks on the human brains, consistently. To alleviate this issue, landmark-matching registration algorithms are proposed by various research groups. Wang et al. \cite{Wang05,Luilandmark,Lui,Lui10,Joshi,Lin} proposed to compute the optimized conformal parameterizations of brain surfaces by minimizing a compounded energy \cite{Wang05,Luilandmark}. The obtained registration can obtain an optimized conformal map that better aligns the features, however, landmarks cannot be exactly matched. Besides, bijectivity cannot be ensured when large number of landmark constraints are enforced. To solve this problem, smooth vector field has also been applied to obtain surface registration. Lui et al. \cite{Lui,Lui10} proposed the use of vector fields to represent surface maps and reconstruct them through integral flow equations. They obtained shape-based landmark matching harmonic maps by looking for the best vector fields minimizing a shape energy. The use of vector fields to compute the registration makes optimization easier, although it cannot describe all surface maps. An advantage of this method is that exact landmark matching can be guaranteed. Time dependent vector fields can also be used. For example, Joshi et al. \cite{Joshi} proposed the generation of large deformation diffeomorphisms for landmark point matching, where the registrations are generated as solutions to the transport equation of time dependent vector fields. The time dependent vector fields facilitate the optimization procedure, although it may not be a good representation of surface maps since it requires more memory. The computational cost of the algorithm is also expensive.

\begin{figure*}[t]
\centering
\includegraphics[height=1.5in]{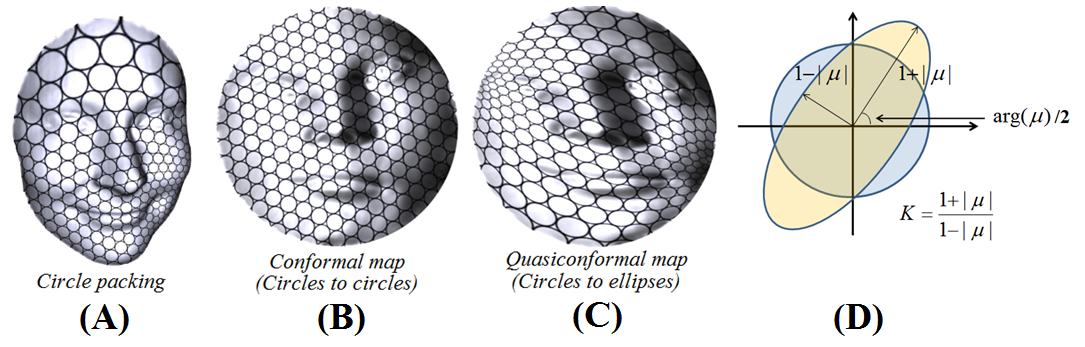}
\caption{(A) shows a human face with circle packing texture. Under the conformal parameterization, infinitesimal circles are mapped to circles as shown in (B). Under quasi-conformal parameterization, infinitesimal circles are mapped to ellipse as shown in (C). (D) illustrates how the Beltrami coefficient measure the conformality distortion of a quasi-conformal map. \label{fig:illustration}}
\end{figure*}
\section{Mathematical Background}

In this section, we describe some basic mathematical concepts related to our algorithms. For details, we refer the readers to \cite{Gardiner,Lehto,DiffGeomBook}.

A surface $S$ with a conformal structure is called a \emph{Riemann surface}. Given two Riemann surfaces $M$ and $N$, a map $f:M\to N$ is \emph{conformal} if it preserves the surface metric up to a multiplicative factor called the conformal factor. An immediate consequence is that every conformal map preserves angles. With the angle-preserving property, a conformal map effectively preserves the local geometry of the surface structure. 

A generalization of conformal maps is the \emph{quasi-conformal} maps, which are orientation preserving homeomorphisms between Riemann surfaces with bounded conformality distortion, in the sense that their first order approximations takes small circles to small ellipses of bounded eccentricity \cite{Gardiner}. Mathematically, $f \colon \mathbb{C} \to \mathbb{C}$ is quasi-conformal provided that it satisfies the Beltrami equation:
\begin{equation}\label{beltramieqt}
\frac{\partial f}{\partial \overline{z}} = \mu(z) \frac{\partial f}{\partial z}.
\end{equation}
\noindent for some complex valued function $\mu$ satisfying $||\mu||_{\infty}< 1$. $\mu$ is called the \emph{Beltrami coefficient}, which is a measure of non-conformality. In particular, the map $f$ is conformal around a small neighborhood of $p$ when $\mu(p) = 0$. Infinitesimally, around a point $p$, $f$ may be expressed with respect to its local parameter as follows:
\begin{equation}
\begin{split}
f(z) & = f(p) + f_{z}(p)z + f_{\overline{z}}(p)\overline{z} \\
& = f(p) + f_{z}(p)(z + \mu(p)\overline{z}).
\end{split}
\end{equation}

\begin{figure*}[t]
\centering
\includegraphics[height=1.5in]{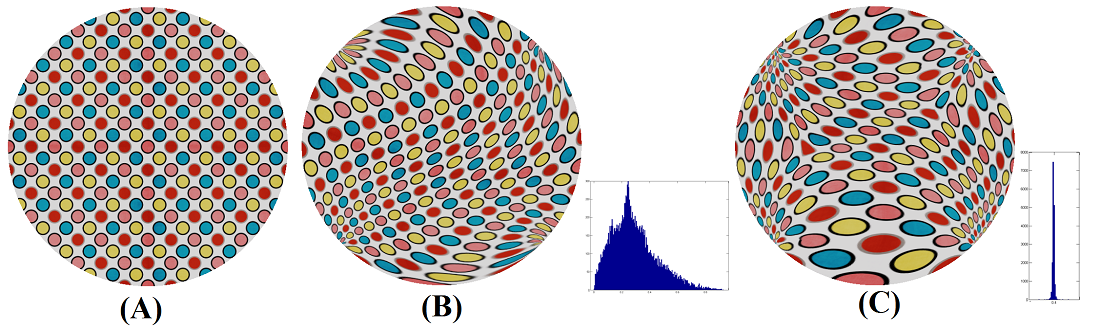}
\caption{Difference between a general QC map and a Teichm\"uller map. (A) shows the original textured mesh. It is mapped to another disk by a general QC map. Note that the distribution of the norm of BC are spread out. (C) shows the Teichm\"uller map, whose BC norm is concentrated near 0.4. \label{fig:qcillustration}}
\end{figure*}

Obviously, $f$ is not conformal if and only if $\mu(p)\neq 0$. Inside the local parameter domain, $f$ may be considered as a map composed of a translation to $f(p)$ together with a stretch map $S(z)=z + \mu(p)\overline{z}$, which is postcomposed by a multiplication of $f_z(p),$ which is conformal. All the conformal distortion of $S(z)$ is caused by $\mu(p)$. $S(z)$ is the map that causes $f$ to map a small circle to a small ellipse. From $\mu(p)$, we can determine the angles of the directions of maximal magnification and shrinking and the amount of them as well. Specifically, the angle of maximal magnification is $\arg(\mu(p))/2$ with magnifying factor $1+|\mu(p)|$; The angle of maximal shrinking is the orthogonal angle $(\arg(\mu(p)) -\pi)/2$ with shrinking factor $1-|\mu(p)|$. Thus, the Beltrami coefficient $\mu$ gives us all the information about the properties of the map (See Figure \ref{fig:qcillustration}(D)).

\begin{figure*}[t]
\centering
\includegraphics[height=2.2in]{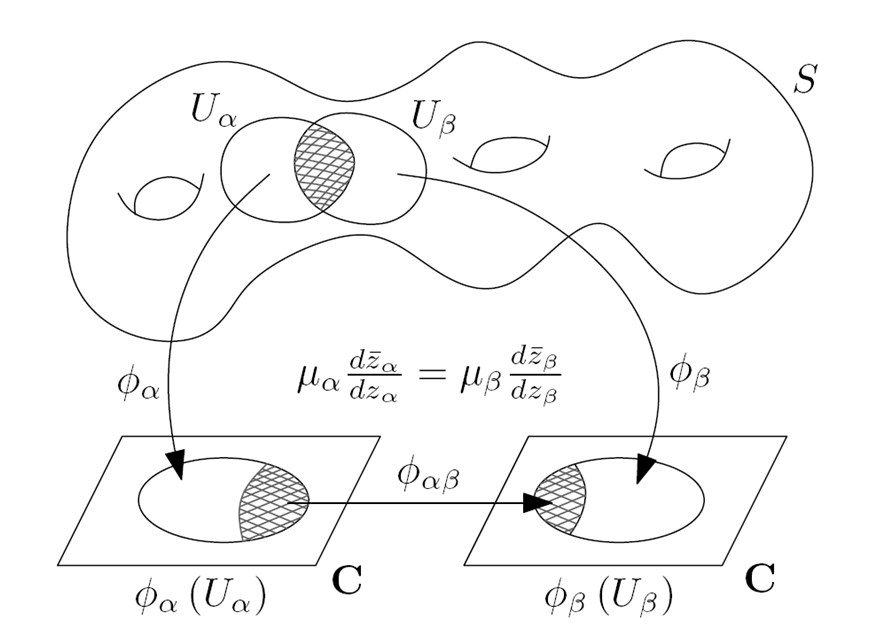}
\caption{Illustration of how Beltrami differential is defined on general Riemann surfaces. \label{fig:chart}}
\end{figure*}

\begin{figure*}[t]
\centering
\includegraphics[height=2.2in]{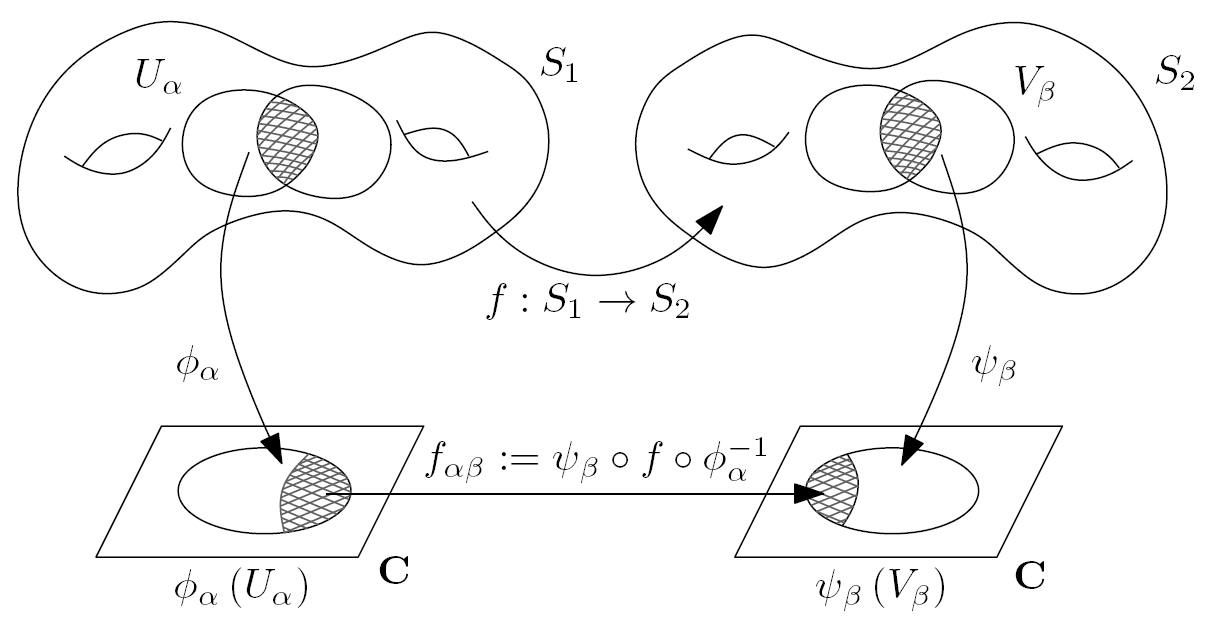}
\caption{Illustration of quasi-conformal mapping between Riemann surfaces. \label{fig:qcillustration2}}
\end{figure*}

The maximal dilation of $\phi$ is given by:
\begin{equation}
K(\phi) = \frac{1+||\mu_{\phi}||_{\infty}}{1-||\mu_{\phi}||_{\infty}}.
\end{equation}

Let $f = u + \sqrt{-1} v$. From the Beltrami equation (\ref{beltramieqt}),
\begin{equation}
\mu(f) = \frac{(u_x - v_y) + \sqrt{-1}\ (v_x + u_y)}{(u_x + v_y) + \sqrt{-1}(v_x - u_y)}
\end{equation}

Let $\mu(f) = \rho + \sqrt{-1}\ \tau $. We can write $v_x$ and $v_y$ as linear combinations of $u_x$ and $u_y$,
\begin{equation}\label{eqt:linearB1cont}
\begin{split}
-v_y & = \alpha_1 u_x + \alpha_2 u_y;\\
v_x & = \alpha_2 u_x + \alpha_3 u_y.
\end{split}
\end{equation}
\noindent where $\alpha_1 = \frac{(\rho -1)^2 + \tau_T^2}{1-\rho^2 - \tau^2} $; $\alpha_2 = -\frac{2\tau}{1-\rho^2 - \tau^2} $; $\alpha_3 = \frac{1+2\rho+\rho^2 +\tau^2}{1-\rho^2 - \tau^2} $.

Similarly,
\begin{equation} \label{eqt:linearB2cont}
\begin{split}
-u_y & = \alpha_1 v_x + \alpha_2 v_y;\\
u_x & = \alpha_2 v_x + \alpha_3 v_y.
\end{split}
\end{equation}

Since $\nabla \cdot \left(\begin{array}{c}
-v_y\\
v_x \end{array}\right) = 0$, we obtain
\begin{equation}\label{eqt:BeltramiPDE}
\nabla \cdot \left(A \left(\begin{array}{c}
u_x\\
u_y \end{array}\right) \right) = 0\ \ \mathrm{and}\ \ \nabla \cdot \left(A \left(\begin{array}{c}
v_x\\
v_y \end{array}\right) \right) = 0
\end{equation}

Quasiconformal mapping between two Riemann surfaces $R_1$ and $R_2$ can also be defined. Instead of the Beltrami coefficient, the {\it Beltrami differential} has to be used. A Beltrami differential $\mu(z) \frac{\overline{dz}}{dz}$ on the Riemann surface $R_1$ is an assignment to each chart $(U_{\alpha},\phi_{\alpha})$ of an $L_{\infty}$ complex-valued function $\mu_{\alpha}$, defined on local parameter $z_{\alpha}$ such that
\begin{equation}
\mu_{\alpha}(z_{\alpha})\frac{d\overline{z_{\alpha}}}{dz_{\alpha}} = \mu_{\beta}(z_{\beta})\frac{d\overline{z_{\beta}}}{dz_{\beta}},
\end{equation}
\noindent on the domain which is also covered by another chart $(U_{\beta},\phi_{\beta})$, where $\frac{dz_{\beta}}{dz_{\alpha}}= \frac{d}{dz_{\alpha}}\phi_{\alpha \beta}$ and $\phi_{\alpha \beta} = \phi_{\beta}\circ \phi_{\alpha}^{-1}$ (See Figure \ref{fig:chart}).

An orientation preserving diffeomorphism $f:R_1 \to R_2$ is called quasi-conformal associated with $\mu(z) \frac{\overline{dz}}{dz}$ if for any chart $(U_{\alpha},\phi_{\alpha})$ on $R_1$ and any chart $(V_{\beta},\psi_{\beta})$ on $R_2$, the mapping $f_{\alpha \beta}:= \psi_{\beta}\circ f\circ f_{\alpha}^{-1}$ is quasi-conformal associated with $\mu_{\alpha}(z_{\alpha})\frac{d\overline{z_{\alpha}}}{dz_{\alpha}}$ (See Figure \ref{fig:qcillustration2}.

Now, suppose $R_1$ and $R_2$ are open Riemann surfaces with the same topology. The boundary dilation $K_1[\phi]$ of $\phi$ is the infimum of the maximal dilation of $\psi|_U$ over all quasi-conformal maps $\psi$ isotopic to $\phi$ relative to the boundary and all neighborhoods $U$ of $\partial R_1$.

In case $R_1$ and $R_2$ are simply-connected, conformal mapping between $R_1$ and $R_2$ always exists. However, conformal mapping may not exist between multiply-connected domains. For example, there is generally no conformal mapping between two annuli with different radii of inner circles. One might be interested in investigating extremal quasiconformal mappings, which is extremal in the sense of minimizing the $||\cdot||_{\infty}$ over all Beltrami differentials corresponding to quasi-conformal mappings between $R_1$ and $R_2$. Extremal mapping always exists but need not be unique. More specifically, an extremal quasi-conformal mapping can be defined mathematically as follows:

\bigskip

\begin{defn}
Let $f: R_1\to R_2$ be a quasi-conformal mapping between $R_1$ and $R_2$. $f$ is said to be an extremal mapping if for any quasi-conformal mapping $h: R_1\to R_2$ isotopic to $\phi$ relative to the boundary,
\begin{equation}\label{inequality}
K(\phi) \leq K(\psi)
\end{equation}
\noindent It is uniquely extremal if the inequality (\ref{inequality}) is strict.
\end{defn}

\bigskip

Another kind of mapping, called the {\it Teichm\"uller mapping}, is closely related to extremal mapping. Teichm\"uller mapping is defined as follows:

\bigskip

\begin{defn}
Let $f:R_1\to R_2$ be a quasi-conformal mapping. $f$ is said to be a Teichm\"uller mapping associated with $\varphi:R_1\to \mathbb{C}$ if its associated Beltrami coefficient is of the form:
\begin{equation}\label{Teichmullermap}
\mu(f) = k \frac{\overline{\varphi}}{|\varphi|}
\end{equation}
\noindent for some constant $k <1$ and $\varphi \neq 0$.
\end{defn}

\bigskip

Extremal mapping is not unique for general cases. However, a Teichm\"uller mapping associated with a holomorphic function is the unique extremal mapping in its homotopic class. The Strebel's theorem explains the relationship bewtween Teichm\"uller mapping and extremal mapping.

\bigskip

\begin{thm}[Strebel's theorem]
Let $f$ be an extremal quasi-conformal mapping with $K(f) >1$. If $K_1[f]< K(f)$, then $f$ is a Teichmuller map associated with an integrable holomorphic quadratic function on $R_1$. Hence, $f$ is also an unique extremal mapping.
\end{thm}

\bigskip

In particular, Teichm\"uller mapping and extremal mapping of the unit disk are closely related.

\bigskip

\begin{thm}
Let $g:\partial \mathbb{D} \to \partial \mathbb{D}$ be an orientation-preserving homeomorphism of $\partial \mathbb{D}$. Suppose further that $h'(e^{i\theta}) \neq 0$ and $h''(e^{i\theta})$ is bounded. Then there is a Teichm\"uller mapping $f$ of finite norm that is the uniquely extremal extension of $g$ to $\mathbb{D}$. That is, $f:\mathbb{D}\to \mathbb{D}$ is an extremal mapping with $f|_{\partial \mathbb{D}} = g$.
\end{thm}

\bigskip

In other words, an extremal mapping of the unit disk with suitable boundary condition is a Teichm\"uller mapping. It can thus be obtained by searching for an optimal Beltrami coefficient whose maximal dilatation is the minimum, while its norm is constant everywhere. It turns out that in most situations, extremal quasi-conformal mapping is a Teichm\"uller mapping (even for domains with non-trivial topologies). In some rare situations when an extremal mapping is not exactly a Teichm\"uller mapping, one can get a Teichm\"uller mapping whose dilation is arbitrarily close to the extremal dilation.

\bigskip

\begin{thm}
Let $\mathfrak{F}$ be a class of quasi-conformal mappings between the open Riemann surfaces $R_1$ and $R_2$, which are homotopic modulo the boundary. Let $K_0$ be the smallest maximal dilation of the mappings in $\mathfrak{F}$. Then there are Teichm\"uller mappings in $\mathfrak{F}$, associated with a meromorphic function with at most one simple pole, whose dilation is arbitrarily close to $K_0$.
\end{thm}

\bigskip

\section{Mathematical Formulation}
In this section, we give the mathematical formulation for obtaining the Teichm\"uller extremal mapping with least conformality distortion. We propose to use the Beltrami coefficient(BC) to represent the mapping, instead of the commonly used representations by deformation fields or coordinate functions. The diffeomorphic property of the registration can then be effectively controlled. Our goal is to formulate the problem into a variational problem to obtain an optimal BC, $\mu(f)$, associated to the desired extremal mapping $f$.

Suppose $D_1$ and $D_2$ are two domains in the complex plane with the same topology. $D_1$ and $D_2$ can either be simply-connected or multiply-connected. Suppose the boundary condition of the desired extremal mapping $f:D_1\to D_2$ is known. Denote it by $f|_{\partial D_1}: \partial D_1 \to \partial D_2 = g$. The Teichm\"uller extremal mapping can be mathematically described as follows:

\begin{equation}\label{original}
\frac{\partial f}{\partial \overline{z}} = k\frac{\overline{\varphi}}{|\varphi|} \frac{\partial f}{\partial z} \ \ \mathrm{and\ \ } f|_{\partial D_1} = g \ \ \mathrm{on\ } \partial D_1
\end{equation}

\noindent for some constant $k$ and holomorphic function $\varphi:D_1 \to \mathbb{C}$.

\bigskip

Recall that a Teichm\"uller extremal mapping is extremal in the sense of minimizing the $||\cdot||_{\infty}$ over all Beltrami differentials corresponding to quasiconformal mappings in the Teichm\"uller equaivalence class. In other words, for any $h:D_1 \to D_2$ satisfying $h|_{\partial D_1} = g$, we have
\begin{equation}\label{originalextremal}
||\mu(f)||_{\infty} \leq ||\mu(h)||_{\infty}
\end{equation}
\noindent where $\mu(f)$ and $\mu(h)$ are the Beltrami coefficient of $f$ and $h$ respectively. Hence, our original problem (\ref{original}) can be formulated as a variational problem as follows:
\begin{equation}\label{variational1}
\begin{split}
f & = \mathbf{argmin}_{f:D_1\to D_2} E_1 (f)\\
& := \mathbf{argmin}_{f:D_1\to D_2} \{ ||\mu(f)||_{\infty} + ||\nabla\ |\mu(f)|\ ||_2^2 \}
\end{split}
\end{equation}
\noindent subject to:

\smallskip

\begin{itemize}
\item $f|_{\partial D_1} = g$ (boundary condition);
\item $\mu(f) = k \frac{\overline{\varphi}}{\varphi}$ for some constant $k$ and holomorphic function $\varphi: D_1 \to \mathbb{C}$.
\end{itemize}

\bigskip

Theoretically, a diffeomorphism $f$ is associated to a unique smooth BC $\mu(f)$ with $||\mu(f)||_{\infty} <1$. The Beltrami coefficient $\mu(f)$ measures the conformality distortion of the map $f$. It can be considered as a unique representation of $f$. The first term of the energy functional $E_1$ aims to minimize the maximal conformality distortion of the mapping. The second term minimizes the harmonic energy of $|\mu(f)|$. Since a Teichm\"uller map has a constant norm, minimizing the second term aims to obtain a optimal BC whose norm is constant.


However, minimizing $E_1(f)$ with respect to the space of diffeomorphisms between $D_1$ and $D_2$ is difficult.

\begin{equation}
\begin{split}
f & = \mathbf{argmin}_{f} \{ ||\mu(f)||_{\infty} + ||\nabla\ |\mu(f)|\ ||_2^2 \}\\
& = \mathbf{argmin}_{f} ||\frac{\partial f/\partial \overline{z}}{\partial f/\partial \overline{z}}||_{\infty}+||\nabla\ |\frac{\partial f/\partial \overline{z}}{\partial f/\partial \overline{z}}|\ ||_2^2 \\
\end{split}
\end{equation}
\noindent subject to $f|_{\partial D_1} = g$ and $\mu(f) = k \frac{\overline{\varphi}}{\varphi}$ for some constant $k$ and holomorphic function $\varphi: D_1 \to \mathbb{C}$.

\bigskip

In order to minimize the above constrained minimization effectively, we propose to reformulate the energy functional with respect to space of all Beltrami coefficients:

\begin{equation}\label{variational2}
\begin{split}
(\nu, f) & = \mathbf{argmin}_{\nu:D_1\to \mathbb{C}} E_2 (\nu)\\
& := \mathbf{argmin}_{\nu:D_1\to \mathbb{C}} \{ ||\nu||_{\infty} + ||\nabla\ |\nu|\ ||_2 \}
\end{split}
\end{equation}
\noindent subject to:

\smallskip

\begin{itemize}
\item $\nu = \mu(f)$ and $||\nu||_{\infty} <1$;
\item $\nu = k \frac{\overline{\varphi}}{\varphi}$ for some constant $k$ and holomorphic function $\varphi: D_1 \to \mathbb{C}$;
\item $f|_{\partial D_1} = g$ (boundary condition).
\end{itemize}

\bigskip

In other words, the minimization problem (\ref{variational1}) is reformulated to be optimized with respect to BCs, which are complex-valued functions defined on $D_1$. Minimizing the energy functional with respect to BCs is advantageous since the diffeomorphic property of the mapping can be easily controlled. Every diffeomorphism is associated to a smooth Beltrami coefficient $\mu(f)$. $\mu(f)$ measures the bijectivity (1-1 and onto) of $f$. In fact, $\mu(f)$ is related to the Jacobian $J(f)$ of $f$ by the following formula:

\begin{equation}
|J(f)|^2 = |\frac{\partial f}{\partial z}|^2 (1- |\mu(f)|^2)
\end{equation}

\noindent Therefore, the map $f$ is bijective if $|\mu(f)|$ is everywhere less than 1. When solving the minimization problem (\ref{variational2}), the bijectivity of the mapping in each iterations can be ensured by enforcing $||\nu||_{\infty} <1$. Our goal is to look for an optimized smooth BC, $\nu$, such that its associated quasi-conformal map is our desired Teichm\"uller extremal mapping.

The boundary condition in the variational problem (\ref{variational2}) can be relaxed. The Dirichlet condition defined on the whole boundary is not required. Also, interior landmark constraints can be enforced. Our goal is to solve the variational problem with these landmark constraints, which determine the optimal 1-1 correspondence (including the boundary correspondence) automatically. In other words, the boundary condition in the problem (\ref{variational2}) can be reformulated as:

\begin{equation}
f(a_i)= b_i;\ f(p_j) = q_j;\ \mathrm{for\ } i=1,...,n;\ j=1,...,m
\end{equation}
\noindent where $a_i$ and $b_i$ are corresponding landmark points or curves defined on $\partial D_1$ and $\partial D_2$ respectively; and $p_j$ and $q_j$ are corresponding interior landmark points of curves in $D_1$ and $D_2$ respectively. By optimizing the energy functional \ref{variational2}, an {\it optimized Teichm\"uller extremal mapping} can be obtained, which matches landmark features consistently while minimizing the maximal conformality distortion.

Note also that the above formulation is designed for extremal mapping of 2D connected domains. However, it can easily be extended to simply-connected or multiply-connected open surfaces. Let $S_1$ and $S_2$ be two connected open surfaces with the same topology. We can conformally parameterize $S_1$ and $S_2$ by $\phi_1:S_1 \to D_1 \subset\mathbb{C}$ and $\phi_2: S_2 \to D_2 \subset\mathbb{C}$ respectively. Then the extremal mapping $f:S_1 \to S_2$ between $S_1$ and $S_2$ induces an extremal mapping $\tilde{f}:= \phi_2 \circ f\circ \phi_1^{-1}: D_1\to D_2$. All the above formulation applies to $\tilde{f}$. In other words, the computation of the extremal mapping between connected surfaces embedded in $\mathbb{R}^3$ can be reduced to the computation of the extremal mapping between the conformal domains in $\mathbb{C}$.

In the subsequent section, we propose an algorithm, called the {\it quasi-conformal (QC) iteration} to solve the above minimization problems (\ref{variational2}).

\section{Main Algorithm}
In this section, we describe an iterative scheme, called the {\it quasi-conformal (QC) iteration}, for solving the variational problem (\ref{variational2}). The QC iteration is based on the Linear Beltrami Solver(LBS). The LBS will firstly be explained in detail. QC iteration will then be described.

Practically speaking, 2D domains or surfaces in $\mathbb{R}^3$ are usually represented discretely by triangular meshes. Suppose $K_1$ and $K_2$ are two surface meshes with the same topology representing $S_1$ and $S_2$. We define the set of vertices on $K_1$ and $K_2$ by $V^1 = \{v_i^1\}_{i=1}^n$ and $V^2 = \{v_i^2\}_{i=1}^n$ respectively. Similarly, we define the set of triangular faces on $K_1$ and $K_2$ by $F^1 = \{T_j^1\}_{j=1}^m$ and $F^2 = \{T_j^2\}_{j=1}^m$. Our goal is to look for a piecewise linear homeomorphism between $K_1$ and $K_2$ that approximates the Teichm\"uller extremal mapping between $S_1$ and $S_2$.

\subsection{Linear Beltrami Solver}
Our goal is to look for an optimal Beltrami coefficient(BC) associated to the desired Teichm\"uller mapping. Every quasi-conformal mapping is associated to a unique BC. Given a BC, it is important to have an algorithm to reconstruct the associated quasi-conformal homeomorphism.

Suppose $f:K_1 \to K_2$ is an orientation preserving piecewise linear homeomorphism between $K_1$ and $K_2$. We can assume $K_1$ and $K_2$ are both embedded in $\mathbb{R}^2$. In case $K_1$ and $K_2$ are surface meshes in $\mathbb{R}^3$, we first parameterize them conformally by $\phi_1:K_1\to D_1 \subseteq \mathbb{R}^2$ and $\phi_2:K_2\to D_2\subseteq \mathbb{R}^2$. The composition of $f$ with the conformal parameterizations, $\tilde{f}:= \phi_2 \circ f\circ\phi_1^{-1}$, is then an orientation preserving piecewise linear homeomorphism between $D_1$ and $D_2$ embedded in $\mathbb{R}^2$. In this paper, we assume the topology of the surface mesh is either a connected open surface or a genus-0 closed surface. In other words, the conformal domain $D_i$ ($i=1,2$) can either be a 2D rectangle, unit disk, punctual disk or unit sphere.

To compute the quasi-conformal mapping, the key idea is to discretize Equation \ref{eqt:BeltramiPDE} with two linear systems.

Given a map $f=(u+\sqrt{-1}v): K_1 \to K_2$, we can easily compute its associated Beltrami coefficient $\mu_f$, which is a complex-valued function defined on each triangular faces of $K_1$. To compute $\mu_f$, we simply need to approximate the partial derivatives at every face $T$. We denote them by $D_x f(T) = D_x u + \sqrt{-1} D_x v$ and  $D_y f(T) = D_y u + \sqrt{-1} D_y v$ respectively. Note that $f$ is piecewise linear. The restriction of $f$ on each triangular face $T$ can be written as:
\begin{equation}
f|_T (x,y) = \left( \begin{array}{c}
a_T x + b_T y + r_T \\
c_T x + d_T y + s_T \end{array} \right)
\end{equation}

Hence, $D_x u(T) = a_T$, $D_y u(T) = b_T$,  $D_x v(T) = c_T$ and $D_y v(T) = d_T$. Now, the gradient $\nabla _T f := (D_x f(T), D_y f(T))^t$ on each face $T$ can be computed by solving the linear system:
\begin{equation}\label{eqt:gradient}
\left( \begin{array}{c}
\vec{v}_1 - \vec{v}_0\\
\vec{v}_2 - \vec{v}_0\end{array} \right)\nabla_T \tilde{f}_i = \left( \begin{array}{c}
\frac{\tilde{f}_i(\vec{v}_1) - \tilde{f}_i(\vec{v}_0)}{|\vec{v}_1 - \vec{v}_0|}\\
\frac{\tilde{f}_i(\vec{v}_2) - \tilde{f}_i(\vec{v}_0)}{|\vec{v}_2 - \vec{v}_0|}\end{array} \right),
\end{equation}
\noindent where $[\vec{v_0},\vec{v_1}]$ and $[\vec{v_0},\vec{v_2}]$ are two edges on $T$. By solving equation \ref{eqt:gradient}, $a_T$, $b_T$, $c_T$ and $d_T$ can be obtained. The Beltrami coefficient $\mu_f(T)$ of the triangular face $T$ can then be computed from the Beltrami equation \ref{beltramieqt} by:
\begin{equation}\label{eqt:BC}
\mu_f(T) = \frac{(a_T - d_T)+\sqrt{-1}(c_T + b_T)}{(a_T + d_T)+\sqrt{-1}(c_T - b_T)},
\end{equation}

Equation \ref{eqt:linearB1cont} and \ref{eqt:linearB2cont} are both satisfied on every triangular faces. Let $\mu_f(T) = \rho_T + \sqrt{-1}\ \tau_T$. The discrete versions of Equation \ref{eqt:linearB1cont} and \ref{eqt:linearB2cont} can be obtained.
\begin{equation}\label{eqt:BCsplit1}
\begin{split}
-d_T & = \alpha_1(T) a_T + \alpha_2(T) b_T\\
 c_T & = \alpha_2(T) a_T + \alpha_3(T) b_T
\end{split}
\end{equation}

and

\begin{equation}\label{eqt:BCsplit2}
\begin{split}
-b_T & = \alpha_1(T) c_T + \alpha_2(T) d_T\\
 a_T & = \alpha_2(T) c_T + \alpha_3(T) d_T
\end{split}
\end{equation}
\noindent where:
$\alpha_1(T) = \frac{(\rho_T -1)^2 + \tau_T^2}{1-\rho_T^2 - \tau_T^2} $; $\alpha_2(T) = -\frac{2\tau_T}{1-\rho_T^2 - \tau_T^2} $; $\alpha_3(T) = \frac{1+2\rho_T+\rho_T^2 +\tau_T^2}{1-\rho_T^2 - \tau_T^2} $.

In order to discretize Equation \ref{eqt:BeltramiPDE}, we need to introduce the discrete divergence. The discrete divergence can be defined as follows. Let $T = [v_i,v_j, v_k]$ and $w_I = f(v_I)$ where $I=i,j$ or $k$. Suppose $v_I = g_I + \sqrt{-1}\ h_I$ and $w_I = s_I + \sqrt{-1}\ t_I$ ($I=i,j,k$). Using equation \ref{eqt:gradient}, $a_T, b_T, c_T$ and $d_T$ can be written as follows:
\begin{equation}
\begin{split}
a_T = A_i^T s_i + A_j^T s_j + A_k^T s_k;\ b_T = B_i^T s_i + B_j^T s_j + B_k^T s_k;\\
c_T = A_i^T t_i + A_j^T t_j + A_k^T t_k;\ d_T = B_i^T t_i + B_j^T t_j + B_k^T t_k;
\end{split}
\end{equation}
\noindent where:
\begin{equation}
\begin{split}
&A_i^T = (h_j-h_k )/Area(T),\ A_j^T = (h_k-h_i )/Area(T),\ A_k^T = (h_i-h_j )/Area(T);\\
&B_i^T = (g_k-g_j )/Area(T),\ B_j^T = (g_i-g_k )/Area(T),\ B_k^T = (g_j-g_i )/Area(T);
\end{split}
\end{equation}

Suppose $\vec{V} = (V_1, V_2)$ is a discrete vector field defined on every triangular faces. For each vertex $v_i$, let $N_i$ be the collection of neighborhood faces attached to $v_i$. We define the discrete divergence $Div $ of $\vec{V}$ as follows:
\begin{equation}
Div  (\vec{V})(v_i) = \sum_{T\in N_i} A_i^T V_1(T) + B_i^T V_2(T)
\end{equation}

By careful checking, one can prove that
\begin{equation}
\sum_{T\in N_i} A_i^T b_T = \sum_{T\in N_i} B_i^T a_T;\ \sum_{T\in N_i} A_i^T d_T = \sum_{T\in N_i} B_i^T c_T.
\end{equation}

This gives,
\begin{equation}
Div\ \left(\begin{array}{c}
-D_y u\\
D_x u \end{array}\right) = 0 \ \ \mathrm{and}\ \ Div \left(\begin{array}{c}
-D_y v\\
D_x v\end{array}\right) = 0
\end{equation}

As a result, Equation (\ref{eqt:BeltramiPDE}) can be discretized:
\begin{equation}\label{eqt:BeltramiPDEdiscrete}
Div \left( A \left(\begin{array}{c}
D_x u\\
D_y u\end{array}\right) \right) = 0\ \ \mathrm{and}\ \ Div \left(A \left(\begin{array}{c}
D_x v\\
D_y v\end{array}\right) \right) = 0
\end{equation}
\noindent where $A = \left(\begin{array}{cc}
\alpha_1 & \alpha_2\\
\alpha_2 & \alpha_3 \end{array}\right) $. This is equivalent to:

\begin{equation}\label{eqt:linearB1}
\sum_{T\in N_i} A_i^T [\alpha_1(T) a_T + \alpha_2(T) b_T] + B_i^T[\alpha_2(T) a_T + \alpha_3(T) b_T] = 0
\end{equation}

\begin{equation}\label{eqt:linearB2}
\sum_{T\in N_i} A_i^T [\alpha_1(T) c_T + \alpha_2(T) d_T] + B_i^T[\alpha_2(T) c_T + \alpha_3(T) d_T] = 0
\end{equation}

\noindent for all vertices $v_i \in D$. Note that $a_T$ and $b_T$ can be written as a linear combination of the x-coordinates of the desired quasi-conformal map $f$. Hence, equation \ref{eqt:linearB1} gives us the linear systems to solve for the x-coordinate function of $f$. Similarly, $c_T$ and $d_T$ can also be written as a linear combination of the y-coordinates of the desired quasi-conformal map $f$. Therefore, equation \ref{eqt:linearB2} gives us the linear systems to solve for the y-coordinate function of $f$.

Besides, $f$ has to satisfy certain constraints on the boundary. One common situation is to give the Dirichlet condition on the whole boundary. That is, for any $v_b \in \partial K_1$
\begin{equation}
f(v_b) = w_b \in \partial K_2
\end{equation}

Note that the Dirichlet condition is not required to be enforced on the whole boundary. The proposed algorithm also allows free boundary condition. In the case that $K_1$ and $K_2$ are rectangles, the desired quasi-conformal map should satisfy
\begin{equation}\label{eqt:boundary1}
\begin{split}
f(0) = 0; f(1) = 1\ f(i) = i\ f(1+i) = 1+i;\\
\mathbf{Re}(f) = 0 \mathrm{\ on\ arc\ }[0, i];\ \mathbf{Re}(f) = 1 \mathrm{\ on\ arc\ }[1, 1+i];\\
\mathbf{Imag}(f) = 0 \mathrm{\ on\ arc\ }[0, 1];\ \mathbf{Imag}(f) = 1 \mathrm{\ on\ arc\ }[i, 1+i]
\end{split}
\end{equation}

\begin{figure*}[t]
\centering
\includegraphics[height=2in]{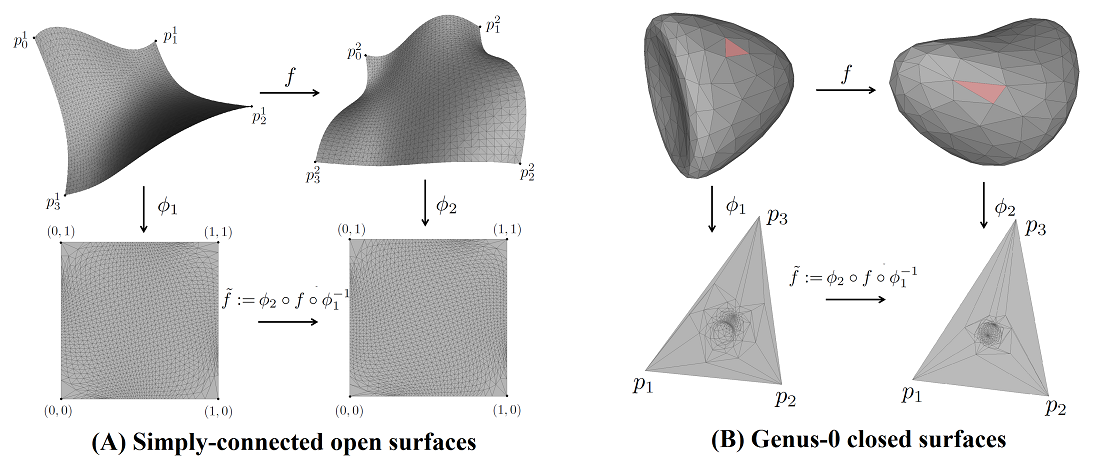}
\caption{Illustration of how Beltrami representations for homeomorphisms between meshes can be computed. (A) shows the case of a homeomorphism between simply-connected open meshes. The two meshes are mapped to a unit square by harmonic parameterizations. (B) shows the case of a homeomorphism between genus-0 closed surface meshes. The two meshes are parameterized onto a triangle in R2, after cutting away a triangular face on each mesh. \label{fig:parameterization}}
\end{figure*}

When $K_i$ ($i=1,2$) is an unit disk, we can parameterize it onto a domain $D_i$, which is a triangle with boundary vertices $p_0^i$, $p_1^i$ and $p_2^i$. $p_0^i$ is on the y-axis whereas  $p_1^i$ and $p_2^i$ are on the x-axis. This can be done by removing a triangular face at the point 1 and map $K_i$ to the upper half plane using a Mobi\"us transformation: $\psi(z) = \sqrt{-1}\frac{1+z}{1-z}$. In this case, the desired quasi-conformal map $f$ should satisfy
\begin{equation}\label{eqt:bounday2}
f(p_0^1) = p_0^2; f(p_1^1) = p_1^2\ \ \mathrm{and\ \ }\mathbf{Imag}(f) = 0 \mathrm{\ on\ arc\ }[p_0^1, p_1^1];
\end{equation}

When $K_i$ ($i=1,2$) is a genus-0 closed surface mesh, we can again parameterize it onto a domain $D_i$, which is a triangle with boundary vertices $p_0^i$, $p_1^i$ and $p_2^i$. This can be done by removing a triangular face at the north pole and map $K_i$ to the 2D plane using stereographic projection. In this case, the desired quasi-conformal map $\tilde{f}$ should satisfy
\begin{equation}\label{eqt:bounday3}
f(p_0^1) = p_0^2; f(p_1^1) = p_1^2\ \ \mathrm{and\ \ }f(p_2^1) = p_2^2
\end{equation}

Suppose landmark correspondences $\{p_i\}_{i=1}^n \leftrightarrow \{q_i\}_{i=1}^n$ are enforced, one should add this constraint to the linear system. Mathematically, it is described as $f(p_i)= q_i$ ($i=1,2,...,n$).

Equations \ref{eqt:linearB1} and \ref{eqt:linearB2} together with the above boundary conditions give a non-singular linear system to solve for $f$. The linear system is symmetric positive definite. Hence, it can be solved effectively by the conjugate gradient method. We call this algorithm the {\it Linear Beltrami Solver}(LBS). Given a Beltrami coefficient $\nu$, we denote the obtained quasi-conformal map from LBS by $\mathbf{LBS}(\nu)$. If landmark constraints are enforced, we denote it by $\mathbf{LBS}_{LM} (\nu)$.

We note that given an arbitrary Beltrami coefficient $\nu$ and arbitrary landmark correspondences, a quasi-conformal mapping associated to $\nu$ might not exist. However, the Linear Beltrami Solver looks for the best quasi-conformal mapping whose Beltrami coefficient closely resemble to $\nu$.

\subsection{Quasi-conformal(QC) iterations}
With the Linear Beltrami Solver, one can easily obtain the best quasi-conformal mapping associated with a given BC. In order to obtain the extremal mapping $f$, our goal is to iteratively search for the optimal BC associated to $f$. With the optimal BC, the desired extremal mapping $f$ can be easily reconstructed using the Linear Beltrami Solver.

Recall that our problem of computing the extremal mapping can be converted into an optimization problem:
\begin{equation}
(\nu, f) = \mathbf{argmin}_{\nu:D_1\to \mathbb{C}} \{ ||\nu||_{\infty} + ||\nabla\nu||_2\}
\end{equation}
\noindent subject to: (1) $\nu = \mu(f)$ and $||\nu||_{\infty} <1$; (2) $\nu = k \frac{\overline{\varphi}}{\varphi}$ for some constant $k$ and holomorphic function $\varphi: D_1 \to \mathbb{C}$; and (3) $f$ satisfies certain boundary condition and/or landmark constraints. Note that the boundary condition in (3) can either be a Dirichlet condition defined on the whole boundary or free boundary condition. In this subsection, we introduce the {\it Quasi-conformal(QC) iteration} to solve the optimization problem.

The QC iteration starts with an initial map $f_0: D_1\to D_2$ satisfying the given boundary condition and landmark constraints. The initial map is chosen to be the quasi-conformal mapping obtained from LBS associated to the BC $\mu_0 = 0$. In other words,
\begin{equation}
f_0 = \mathbf{LBS}_{LM} (\mu_0 :=0)
\end{equation}

Note that with the enforced landmark constraints, the Beltrami coefficient associated to $f_0$ might not be equal to $\nu_0$. The Linear Beltrami Solver simply look for the best quasi-conformal mapping whose Beltrami coefficient resemble to $\nu$ as much as possible. Let $\nu_0$ be the Beltrami coefficient associated to $f_0$. This gives us a pair $(f_0, \nu_0)$, for which $\nu_0 = \mu(f_0)$.

Now, in order to minimize the energy function $E_2$ satisfying condition (2), we propose to perform a Laplace smooth $\mathfrak{L}$ and averaging $\mathfrak{A}$ on $\nu_0$. The Laplace smooth $\mathfrak{L}$, which aims to minimize $E_2$, is given by the following:
\begin{equation}\label{Laplace}
\mathfrak{L}(\nu_0) (T):= \sum_{T_i \in \mathrm{Nbhd}(T)} \nu_0(T) \ /|\mathrm{Nbhd}(T)|
\end{equation}

\noindent where $T$ is a triangular face of $K_1$, $\mathrm{Nbhd}(T)$ is the set of neighborhood faces of $T$ and $|\mathrm{Nbhd}(T)|$ is the number of neighborhood faces in the set $\mathrm{Nbhd}(T)$. Set $\tilde{\mu_1} (T) =  \mathfrak{L}(\nu_0) (T)$.

The averaging operator $\mathcal{A}$ is defined as follows:
\begin{equation}\label{averaging}
\mu_1 (T) =  \mathcal{A}(\tilde{\mu_1}) (T):= (\frac{\sum_{T \in \ \mathrm{all\ faces\ of\ } K_1} |\tilde{\mu_1}|(T)}{\mathrm{No.\ of\ faces\ of\ } K_1}) \frac{\tilde{\mu_1}(T)}{|\tilde{\mu_1}(T)|}
\end{equation}

\noindent $\mathcal{A}$ aims to obtain an optimal $\nu$ satisfying the condition (3) in the optimization problem. An updated quasi-conformal, $f_1$ can then be obtained by LBS: $f_1 = \mathbf{LBS}_{LM} (\mu_1)$. And an updated Beltrami coefficient, $\nu_1: = \mu(f_1)$, can be computed. Thus, we get a new pair $(f_1, \nu_1)$.

The procedure continues until the iteration converges. More specifically, given the pair $(f_n, \nu_n)$ obtained at the $n$ iteration, we can obtain a new pair $(f_{n+1}, \nu_{n+1})$ as follows:
\begin{equation}
\begin{split}
&\mu_{n+1} := \mathcal{A}(\mathfrak{L}(\nu_n));\\
& f_{n+1} := \mathbf{LBS}_{LM} (\mu_{n+1});\\
& \nu_{n+1} := \mu(f_{n+1}).
\end{split}
\end{equation}

Consequently, we get a sequence of pair $(f_n, \nu_n)$, which converges to the optimal Beltrami coefficient associated to the extremal mapping or optimized Teichm\"uller mapping. In practice, we stop the iteration when $||\nu_{n+1} - \nu_n|| < \epsilon$.

We summarize the QC iteration as follows.

\medskip

\noindent $\mathbf{Algorithm\ 5.1:}$ {\it(QC iteration for open surfaces)}\\
\noindent $\mathbf{Input:}$ {\it Triangular meshes: $K_1$ and $K_2$ and the desired boundary condition}\\
\noindent $\mathbf{Output:}$ {\it Optimal Beltrami coefficient $\nu$ and the extremal mapping $f$}\\
\vspace{-3mm}
\begin{enumerate}
\item {\it  Obtain the initial mapping $f_0 = \mathbf{LBS}_{LM} (\mu_0 :=0)$. Set $\nu_0 = \mu(f_0)$;}
\item {\it Given $\nu_n$, compute $\mu_{n+1} := \mathcal{A}(\mathfrak{L}(\nu_n))$; Compute $f_{n+1} := \mathbf{LBS}_{LM} (\mu_{n+1})$ and set $\nu_{n+1} := \mu(f_{n+1})$;}
\item {\it If $||\nu_{n+1} - \nu_n|| \geq \epsilon$, continue. Otherwise, stop the iteration.}
\end{enumerate}

\bigskip

The QC iteration can also be applied to the case when $D_i$ ($i=1,2$) is a unit sphere. In other words, given a set of landmark constraints between the unit sphere, our goal is to look for the Teichm\"uller extremal mapping $f:D_1\to D_2$. However, special attention has to be paid in this case.

Denote the landmark correspondence by $\{p_i\}_{i=1}^n \leftrightarrow \{q_i\}_{i=1}^n$. We can assume that the north pole is fixed. If not, it can also be achieved by a Mobi\"us transformation. The LBS can be applied to unit spheres, by stereographically projecting $D_i$ onto a big triangles in $\mathbb{R}^2$. However, numerical error near the north pole is inevitable. We therefore propose an alternating scheme to fix this problem.

For the initial map, we add the vertices near the north pole $\{n_j\}_{j=1}^m$ ($z > 0.99$) as landmarks and fix all $\{n_j\}_{j=1}^m$. We then compute the Teichm\"uller mapping $f_0$ using Algorithm 5.1. Numerical error will be introduced near the north pole. To fix it, in our next step, we consider the vertices $\{s_j\}_{j=1}^m$ near the south pole ($z<-0.99$) as landmarks. The correspondence is given by: $s_j \leftrightarrow f_1(s_j)$. Rotate the south pole of $D_i$ to the north pole by a Mobi\"us transformation. We can again compute the Teichm\"uller mapping $f_1$ using Algorithm 5.1.

We continue this process until the iteration converges. More specifically, at the $n$ iteration where $n$ is an even integer, we add vertices $\{s_j\}_{j=1}^m$ around south pole as landmarks. Set correspondence as: $s_j \leftrightarrow f_n(s_j)$. Rotate the south pole of $D_i$ to the north pole by a Mobi\"us transformation, and obtain the Teichm\"uller mapping $f_{n+1}$ using Algorithm 5.1. When $n$ is an odd integer, we add vertices $\{n_j\}_{j=1}^m$ around north pole as landmarks. Set correspondence as: $n_j \leftrightarrow f_n(n_j)$ and obtain the Teichm\"uller mapping $f_{n+1}$ using Algorithm 5.1. Set $\nu_{n+1} = \mu(f_n)$.

This alternating process between the north pole and the south pole continues until $||\nu_{n+1}-\nu_n||<\epsilon$.

When $D_i$ ($i=1,2$) is a unit disk, the LBS would also introduce numerical error near 1. To fix it, the same alternating algorithm between 1 and -1 can be applied.

The detailed algorithm can be summarized as follows:

\bigskip

\noindent $\mathbf{Algorithm\ 5.2:}$ {\it(QC iteration for genus-0 closed surfaces)}\\
\noindent $\mathbf{Input:}$ {\it Triangular meshes: $K_1$ and $K_2$ and the desired boundary condition}\\
\noindent $\mathbf{Output:}$ {\it Optimal Beltrami coefficient $\nu$ and the extremal mapping $f$}\\
\vspace{-3mm}
\begin{enumerate}
\item {\it  Add vertices around north pole as landmarks and fix their positions. Obtain the initial Teichm\"uller mapping $f_0$ using Algorithm 5.1. Set $\nu_0 = \mu(f_0)$;}
\item {\it Given $f_n$ and $\nu_n$. When $n$ is even, add vertices $\{s_j\}_{j=1}^m$ around south pole as landmarks. Set correspondence as: $s_j \leftrightarrow f_n(s_j)$. Rotate the south pole of $D_i$ to the north pole. When $n$ is odd, add vertices $\{s_j\}_{j=1}^m$ around south pole as landmarks. Set correspondence as: $s_j \leftrightarrow f_n(s_j)$. Obtain the Teichm\"uller mapping $f_{n+1}$ using Algorithm 5.1. Set $\nu_{n+1} = \mu(f_n)$;;}
\item {\it If $||\nu_{n+1} - \nu_n|| \geq \epsilon$, continue. Otherwise, stop the iteration.}
\end{enumerate}
\begin{figure*}[t]
\centering
\includegraphics[height=3in]{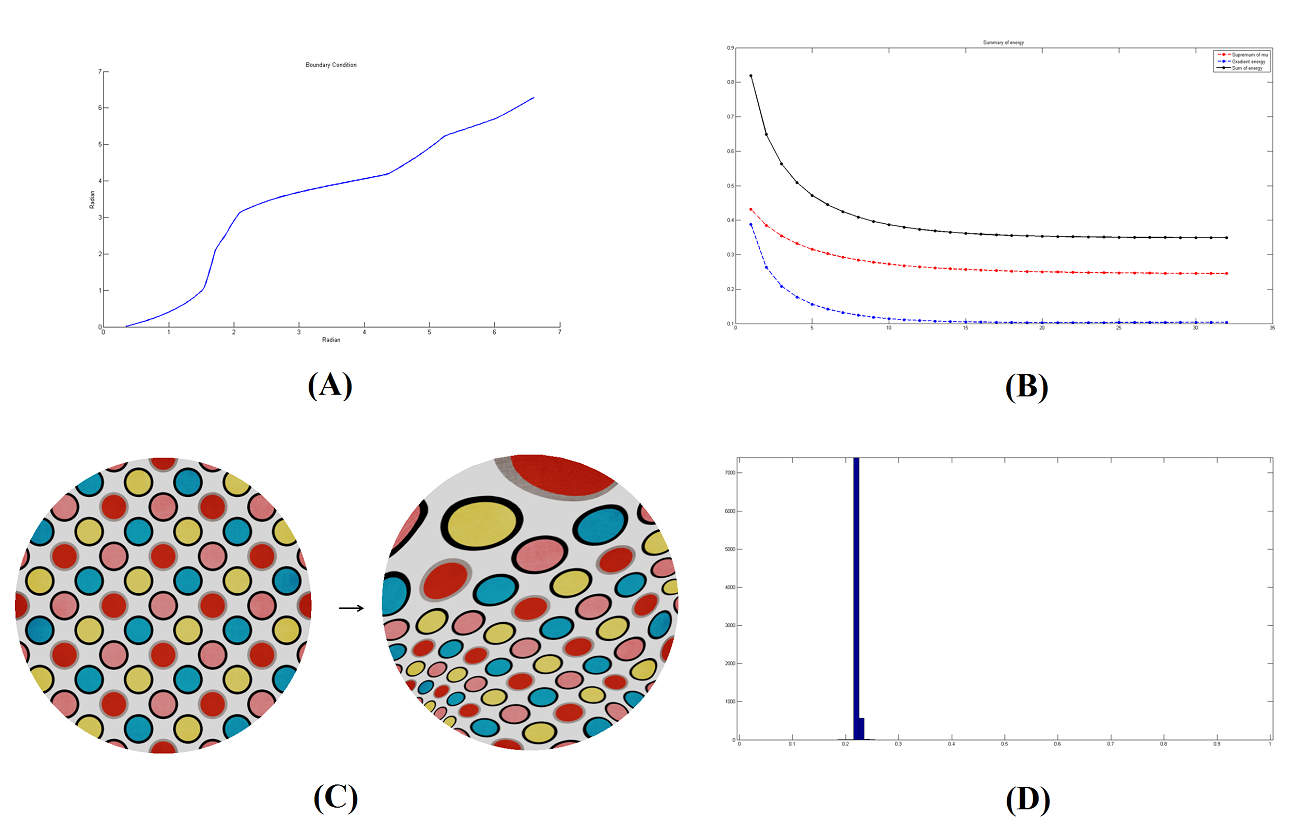}
\caption{Example of the Teichm\"uller extremal mapping of the disk with fixed Dirichlet boundary condition. (A) shows the boundary condition. (B) shows the energy and the sup-norm of the BC in the QC iteration. (C) shows the obtained extremal mapping, visualized using the texture mapping. (D) shows the histogram of the BC norm. \label{fig:Example2}}
\end{figure*}

\begin{figure*}[t]
\centering
\includegraphics[height=1.5in]{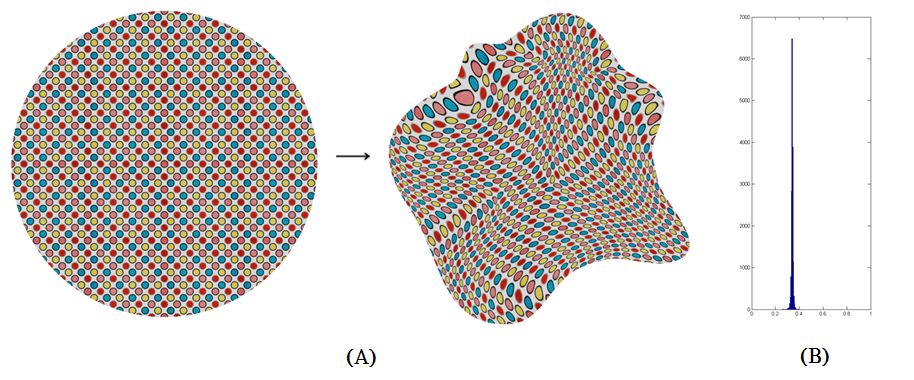}
\caption{Another example of the Teichm\"uller extremal mapping of the disk with fixed Dirichlet boundary condition. (A) shows the obtained extremal mapping, visualized using the texture mapping. (B) shows the histogram of the \label{fig:Example2b}}
\end{figure*}

\begin{figure*}[t]
\centering
\includegraphics[height=2.5in]{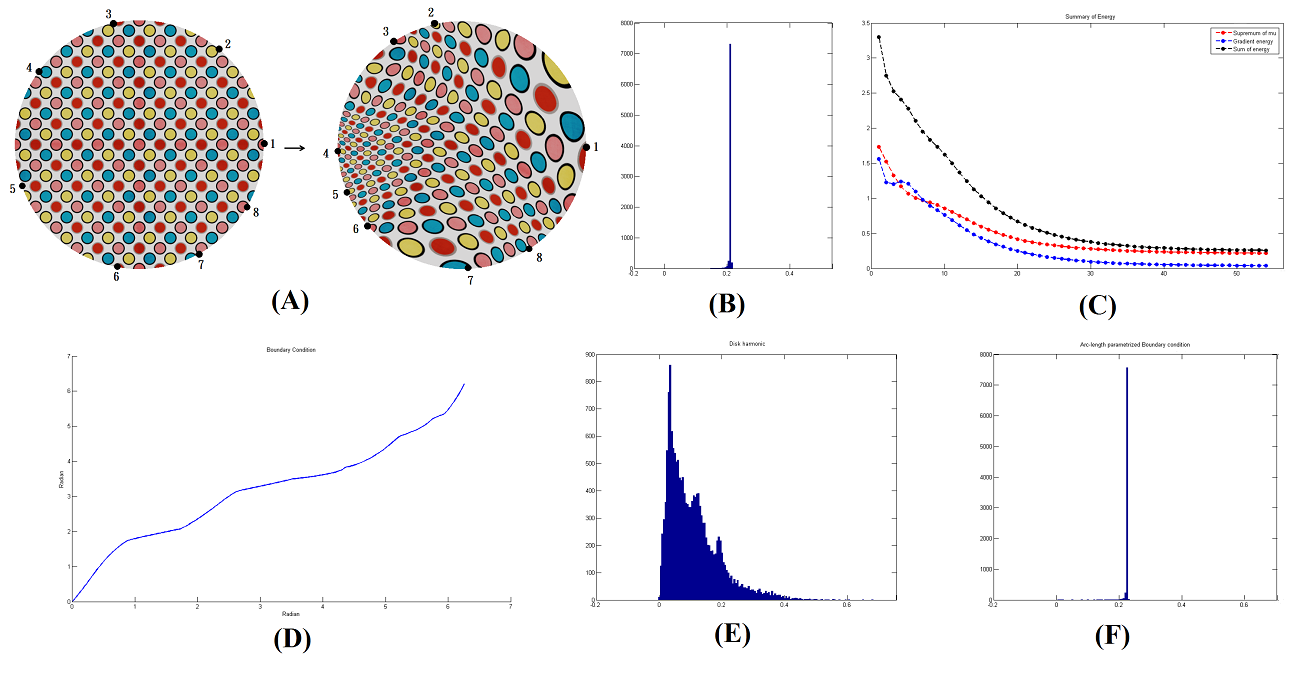}
\caption{Teichm\"uller extremal mapping of the disk with only 8 landmark points constraints on the boundary. (A) shows the Teichm\"uller extremal mapping. (B) shows the histogram of the BC norm. (C) shows the energy and the sup-norm of the BC in the QC iteration. (D) shows the automatically obtained optimal boundary correspondence. (E) shows the histogram of the BC norm under harmonic map with arc-length parameterized boundary condition. (F) shows the histogram of the BC norm under the Teichm\"uller mapping with arc-length parameterized boundary condition.\label{fig:Example4}}
\end{figure*}

\section{Numerical experiments}
In this section, we evaluate our proposed algorithm numerically by synthetic examples.
\subsection{Extremal mapping of simply-connected domains}
In our first numerical experiment, we test our method to compute the extremal mapping of the disk with a given Dirichlet boundary condition. A Dirchlet condition on the whole boundary is given as shown in Figure \ref{fig:Example2}(A). Using the QC iteration, we iteratively obtain the pair $(f_n, \nu_n = \mu(f_n))$. As shown in Figure \ref{fig:Example2}(B), $E_2(f_n,\nu_n)$, $||\nu_n||_{\infty}$ and $||\nabla \nu||_2$ all decrease as iteration increases. The resulting Teichm\"uller extremal mapping is as shown in Figure \ref{fig:Example2}(C), which is visualized using the texture mapping. Note that the original texture is deformed under the extremal mapping. However, the dilations of the ellipses deformed from the small circles are the same. It means the norm of the BC is constant everywhere. The histogram of the norm of the BC is also shown in Figure \ref{fig:Example2}(D), which again demonstrates the norm of the BC is equal to a constant $k=0.15016$. The standard deviation of the BC norm is $0.0034373$. Note that the Dirichlet boundary condition can be of arbitrary shapes. Figure \ref{fig:Example2b} shows the result of finding the Teichm\"uller extremal mapping between the disk and the ameba with given boundary conditions. The resultant norm of BC is also constant everywhere, as demonstrated in Figure \ref{fig:Example2b} (B).
\begin{figure*}[t]
\centering
\includegraphics[height=3in]{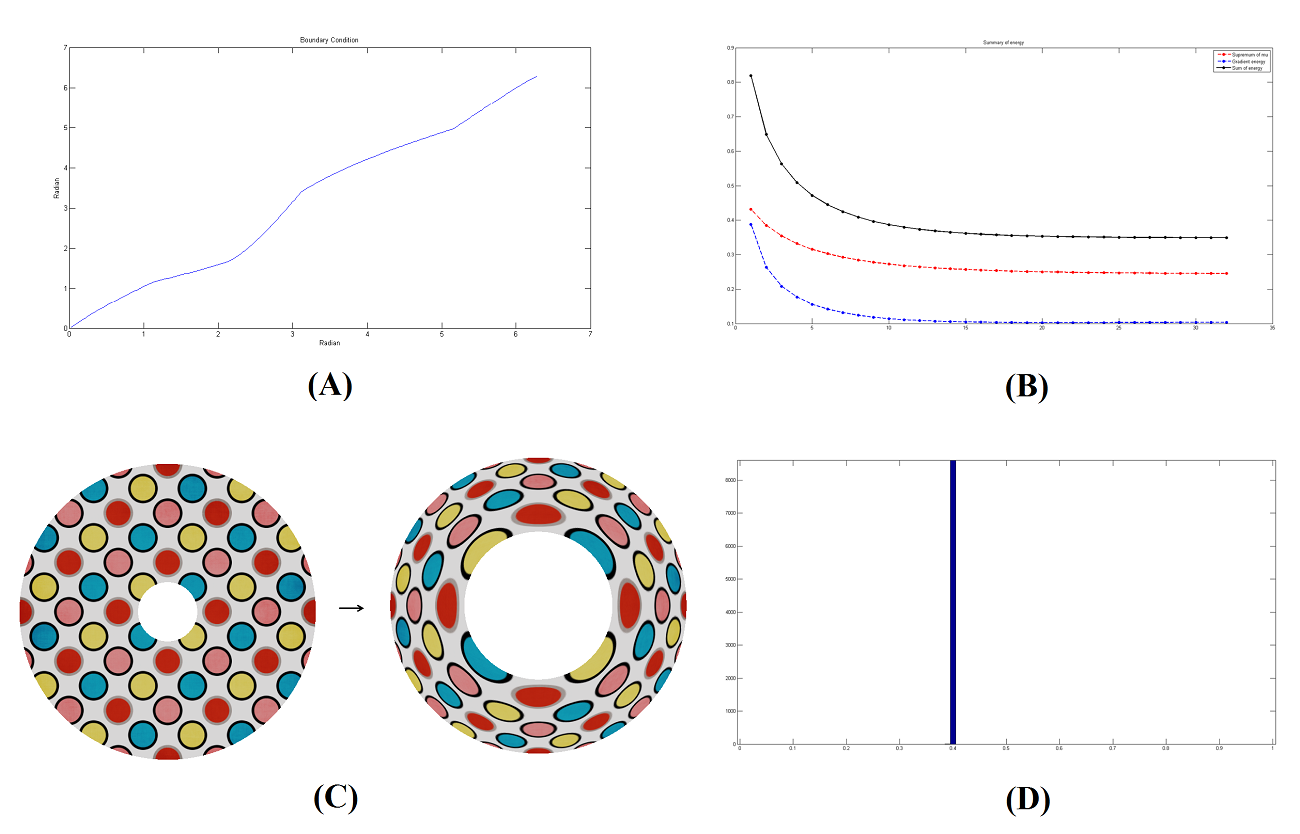}
\caption{Teichm\"uller extremal mapping of the annulus with fixed Dirichlet boundary condition. (A) shows the boundary condition. (B) shows the energy and the sup-norm of the BC in the QC iteration. (C) shows the obtained extremal mapping, visualized using the texture mapping. (D) shows the histogram of the BC norm. \label{fig:Example3}}
\end{figure*}


\begin{figure*}[t]
\centering
\includegraphics[height=1.2in]{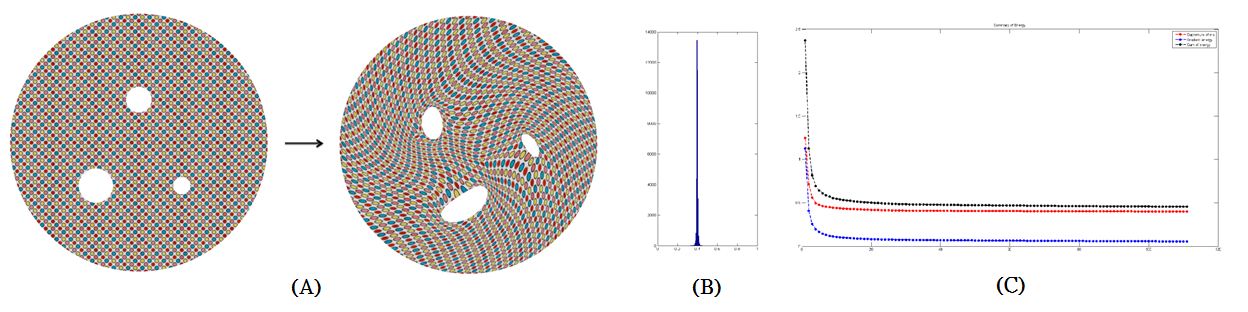}
\caption{Teichm\"uller extremal mapping of the multiply-connected domain containing three holes with fixed Dirichlet boundary condition. (A) shows the Teichm\"uller mapping Dirichlet boundary condition. (B) shows the histogram of the norm of the BC. (C) shows the energy, the sup-norm of the BC and the L2 norm of the gradient of the nom of BC in the QC iteration. \label{fig:3pointsExample}}
\end{figure*}

\begin{figure*}[t]
\centering
\includegraphics[height=1.5in]{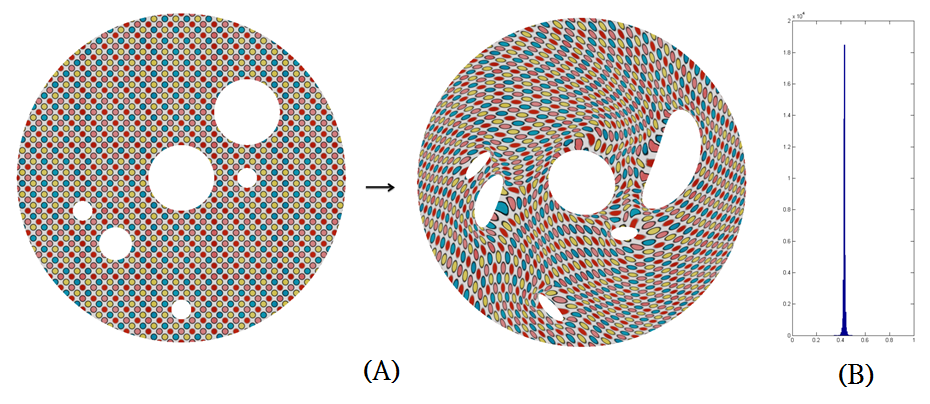}
\caption{Teichm\"uller extremal mapping of the multiply-connected domain containing six holes with fixed Dirichlet boundary condition. (A) shows the Teichm\"uller mapping Dirichlet boundary condition. (B) shows the histogram of the norm of the BC. \label{fig:6pointsExample}}
\end{figure*}

Our algorithm also applies to situation when only a few landmark constraints are enforced on the boundary (instead of the Dirichlet condition defined on the whole boundary). In Figure \ref{fig:Example4}, we test our algorithm to compute Teichm\"uller extremal mapping of the disk with only 8 landmark points constraints on the boundary. (A) shows the the Teichm\"uller extremal mapping. Again, the dilations of the ellipses deformed from the small circles are the same, meaning that the norm of the BC is constant everywhere. (B) shows the histogram of the BC norm. The norm $k$ of the BC is equal to $0.201$. (C) shows the energy and the sup-norm of the BC in the QC iteration. Our algorithm also automatically detect the optimal boundary correspondence. (D) shows the obtained optimal boundary correspondence. (E) shows the histogram of the BC norm under harmonic map with arc-length correspondence on the boundary. Note that the distribution of the conformality distortion is highly non-uniform. (F) shows the histogram of the BC norm under the Teichm\"uller mapping with arc-length parameterized boundary condition (of which the landmark constraints are satisfied). Although a Teichm\"uller mapping can still be obtained, the norm of the BC is equal to 0.23 which is higher than the case when only 8 points landmark constraints are enforced. Hence, the obtained Teichm\"uller mapping is not extremal.


\subsection{Extremal mapping of multiply-connected domains}
Our method can also be applied to multiply-connected domain. In Figure \ref{fig:Example3}, we test our method to compute the extremal mapping of an annulus with Dirichlet boundary condition. The boundary condition is given in (B). Again, the energy functional $E_2$ is decreasing under the QC iteration. The extremal mapping is as shown in (C). The mapping is a Teichm\"uller map, since the BC norm is constant everywhere as shown in (D).

Figure \ref{fig:3pointsExample} and \ref{fig:6pointsExample} show the results of finding the Teichm\"uller extremal mapping with 3 and 6 holes respectively. Figure \ref{fig:3pointsExample} shows the Teichm\"uller map. The corresponding BC norm, sup-norm of BC and the L2 norm of the gradient of the norm of BC are shown in (B) and (C) respectively. Figure \ref{fig:6pointsExample} demonstrates the result of a more complicated multiply-connected domain having 6 holes. These results show that our algorithm can again compute the Teichm\"uller Extremal mapping of multiply-connected domains efficiently and accurately.

\subsection{Optimized Teichm\"uller mapping with interior landmark constraints}
\begin{figure*}[t]
\centering
\includegraphics[height=2.75in]{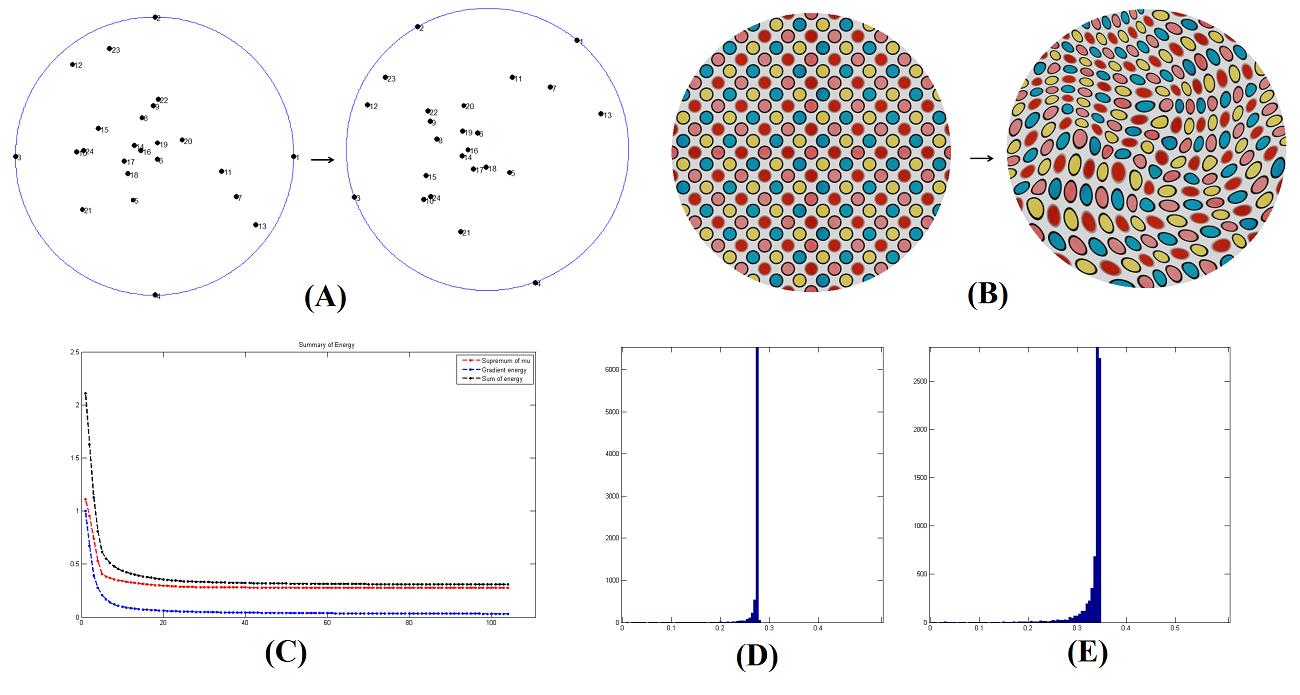}
\caption{Teichm\"uller mapping between the disks with 24 interior landmark constraints enforced. (A) shows the 24 landmark constraints. (B) shows the Teichm\"uller mapping. (C) shows the energy, the sup-norm of the BC and the L-2 norm of the gradient of the BC in the QC iteration. (D) shows the histogram of the norm of the BC. (E) shows the histogram of the BC norm under the Teichm\"uller mapping with arc-length parameterized boundary condition.\label{fig:Example6}}
\end{figure*}

Our algorithm can compute an optimized Teichm\"uller mapping with interior landmark constraints enforced. Figure \ref{fig:Example6} shows the optimized Teichm\"uller mapping between the disk with 25 interior landmark constraints enforced. (A) shows the 24 landmark constraints. (B) shows the obtained Teichm\"uller mapping, visualized using the texture mapping. (C) shows the energy, the sup-norm of the BC and the L-2 norm of the gradient of the BC in the QC iteration. Note that the sup-norm of the BC decreases during the iteration, which illustrates the mapping converges to an optimal mapping minimizing the conformality distortion. (D) shows the histogram of the norm of the BC. The norm $k$ of the BC is uniformly equal to 0.2. (E) shows the histogram of the norm of BC with arc-length boundary correspondence enforced. The norm $k$ of the BC is equal to 0.28, which means the Teichm\"uller mapping is not an extremal one.

\begin{figure*}[t]
\centering
\includegraphics[height=1.35in]{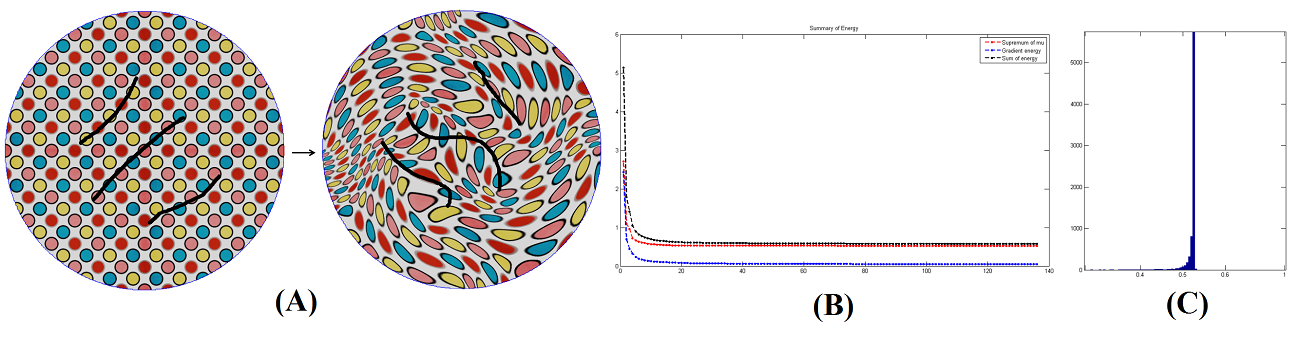}
\caption{Teichm\"uller mapping between the disks with 3 interior landmark curves constraints enforced. (A) shows the Teichm\"uller mapping with 3 landmark curves constraints enforced. (B) shows the energy, the sup-norm of the BC and the L-2 norm of the gradient of the BC in the QC iteration. (C) shows the histogram of the norm of the BC. \label{fig:Example7}}
\end{figure*}

We have test our algorithm to compute the optimized Teichm\"uller mapping of the disk with 3 interior landmark curves constraints enforced in Figure \ref{fig:Example7}. (A) shows the Teichm\"uller mapping with 3 landmark curves constraints enforced. (B) shows the energy, the sup-norm of the BC and the L-2 norm of the gradient of the BC in the QC iteration. The energy is decreasing, indicating that the algorithm converges to an optimized Teichm\"uller mapping. (C) shows the histogram of the norm of the BC. The norm of the BC is accumulated at 0.53.

\begin{figure*}[t]
\centering
\includegraphics[height=1.1in]{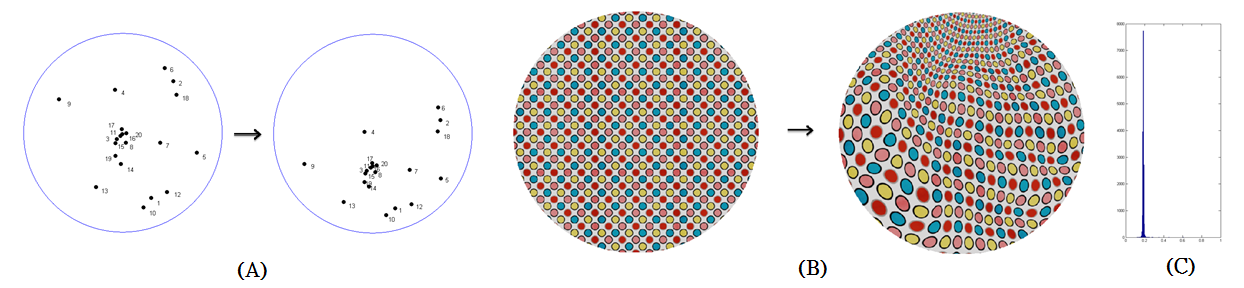}
\caption{Teichm\"uller mapping between the disks with 20 interior landmark constraints enforced. (A) shows the 20 landmark constraints. (B) shows the Teichm\"uller mapping. (C) shows the histogram of the norm of the BC. \label{fig:no_bdy_diskExample}}
\end{figure*}

We also test our algorithm to the case when we only have the correspondence of the interior landmarks. Figure \ref{fig:no_bdy_diskExample} shows an example of finding Teichm\"uller extremal mapping with 20 interior landmark points only. (A) shows the constraint of the feature points. (B) shows the Teichm\"uller mapping. (C) shows the histogram of the norm of the BC. It shows that even boundary condition is not provided, our algorithm can still obtain both the Teichm\"uller extremal mapping which satisfies the landmark constraints and the corresponding optimal boundary condition.

\begin{figure*}[t]
\centering
\includegraphics[height=3in]{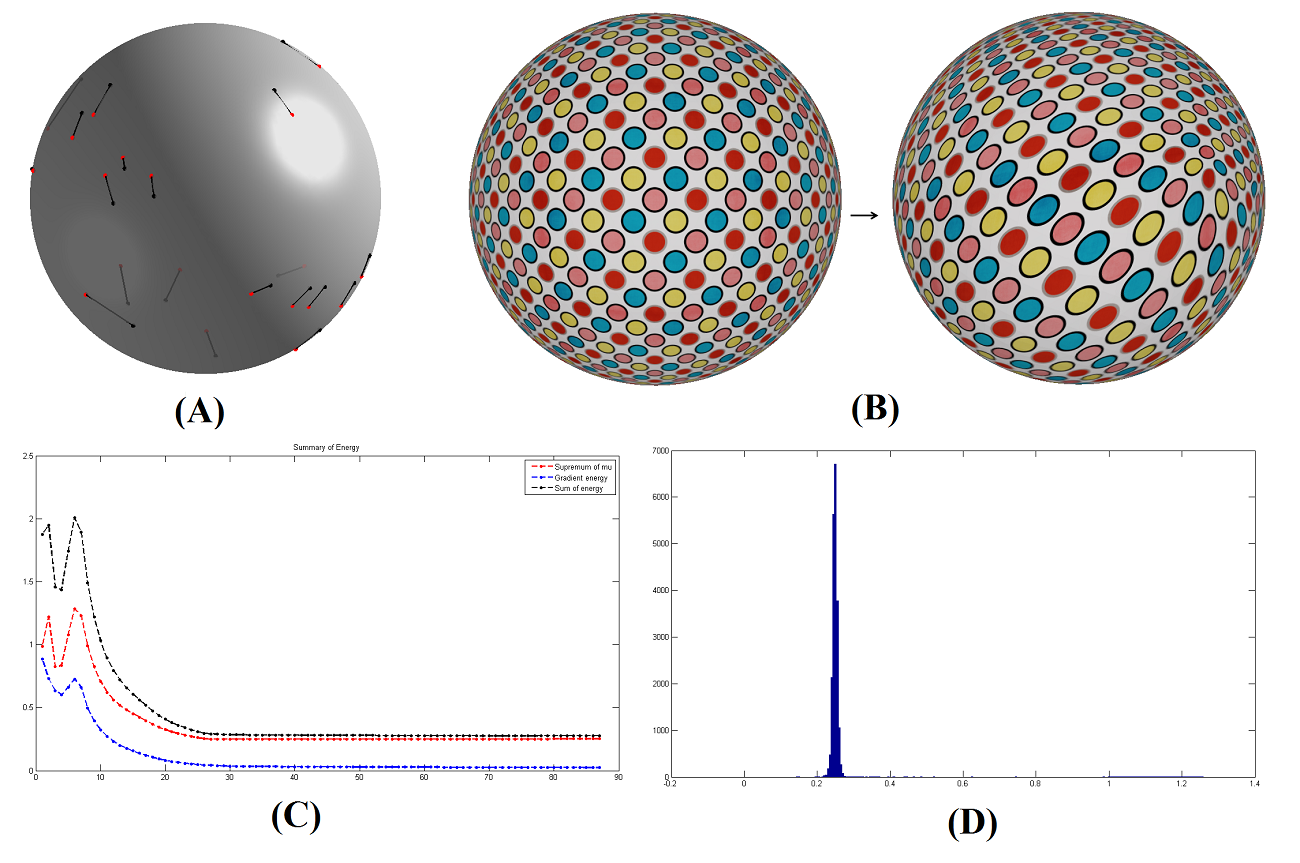}
\caption{Teichm\"uller mapping between the spheres. (A) shows landmark constraints on the sphere. (B) shows the Teichm\"uller mapping. (C) shows the energy, the sup-norm of the BC and the L-2 norm of the gradient of the BC in the QC iteration. (D) shows the histogram of the norm of the BC. \label{fig:sphereExample}}
\end{figure*}

Our algorithm also applies to computing the optimized Teichm\"uller extremal mapping between the unit sphere with interior landmark constraints enforced. Figure \ref{fig:sphereExample} shows an example of Teichm\"uller extremal mapping with 10 interior landmark constraints enforced. (A) shows landmark constraints on the sphere. (B) shows the Teichm\"uller mapping. (C) shows the energy, the sup-norm of the BC and the L-2 norm of the gradient of the BC in the QC iteration. Again, the energy is decreasing, indicating that the algorithm converges to an optimized Teichm\"uller mapping. (D) shows the histogram of the norm of the BC. The norm of the BC is concentrated near 0.21, meaning that the mapping is a Teichm\"uller mapping.

\section{Applications}
In this section, we apply our proposed algorithms for computing landmark matching Teichm\"uller mappings to practical problems. More specifically, we will consider the problems of computing brain landmark matching registrations, constrained texture mappings and human face registrations.

\subsection{Brain landmark matching registration}
\begin{figure*}[t]
\centering
\includegraphics[height=3in]{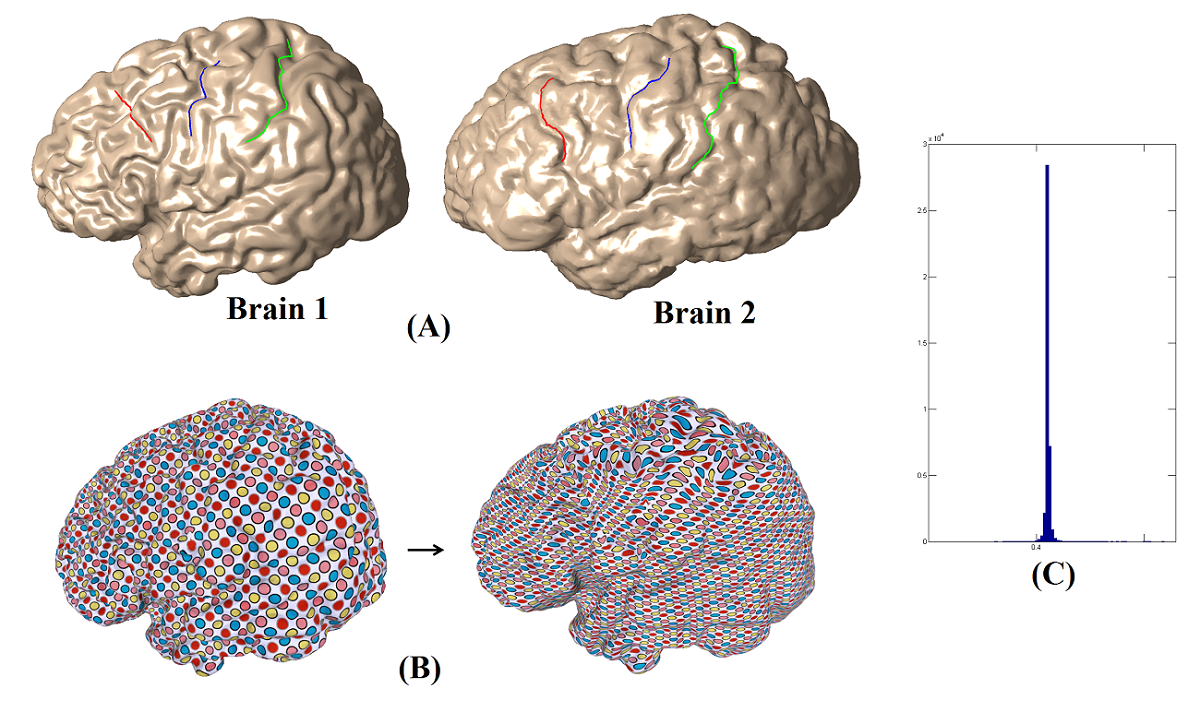}
\caption{(A) shows 2 brain surfaces with 3 corresponding landmarks. (B) shows the Teichm\"uller extremal mapping with 3 landmark constraints enforced. (C) shows the histogram of the norm of BC. \label{fig:brainexample1}}
\end{figure*}

\begin{figure*}[t]
\centering
\includegraphics[height=3in]{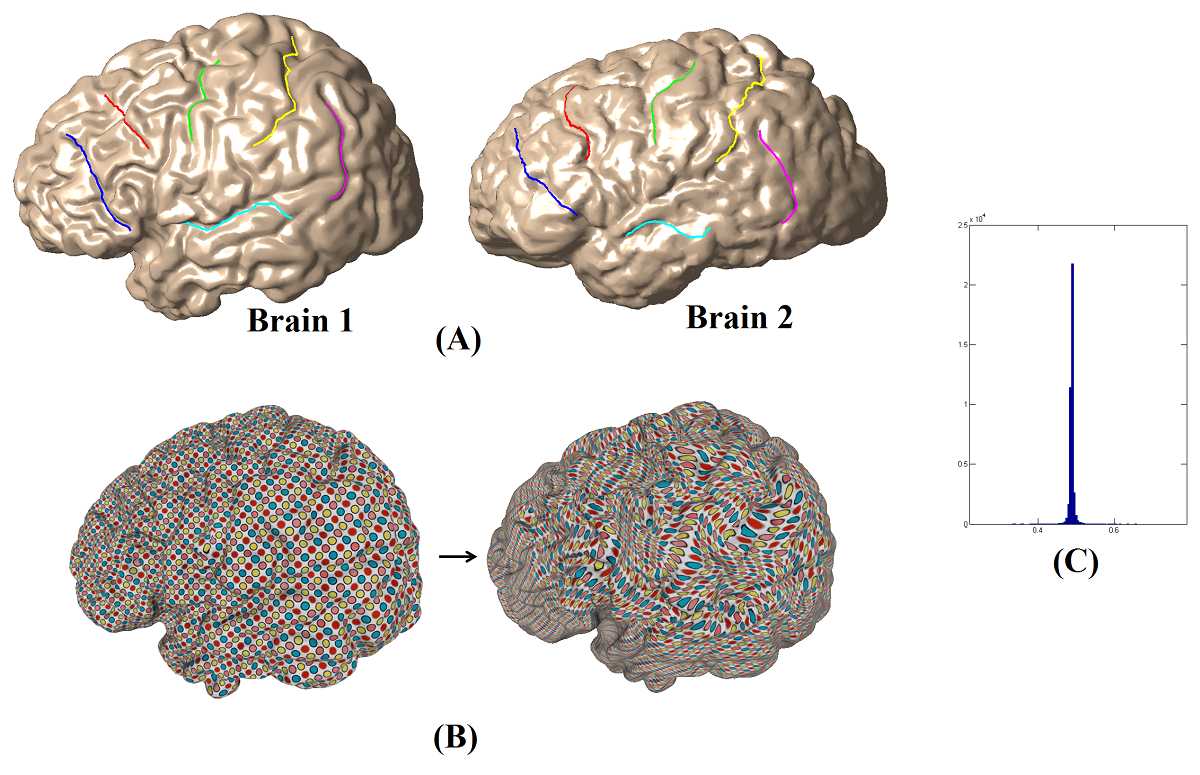}
\caption{(A) shows 2 brain surfaces with 6 corresponding landmarks. (B) shows the Teichm\"uller extremal mapping with 6 landmark constraints enforced. (C) shows the histogram of the norm of BC. \label{fig:brainexample2}}
\end{figure*}

Landmark-based surface registrations are commonly applied for finding meaningful 1-1 correspondences between human brain cortical surfaces. On cortical surfaces, sulcal landmarks can be labeled either manually by neuroscientists or automatically based on various geometric quantities. The sulcal landmarks are important anatomical features. It is therefore desirable to obtain a registration between the cortical surfaces with least geometric distortion, which matches the sulcal landmarks as much as possible. Our algorithms for computing landmark matching Teichm\"uller mappings can be applied. In Figure \ref{fig:brainexample1}, we apply our algorithm to compute the Teichm\"uller mapping between 2 different brain surfaces with 3 corresponding landmarks labeled. (A) shows the corresponding sulcal landmarks, indicated by different colors. (B) shows the obtained Teichm\"uller extremal mapping with 3 landmark constraints enforced, visualized by the circle packing textures. The sulcal landmarks are exactly matched under the mapping. (C) shows the histogram of the norm of the associated BC. The norm is a constant showing that the obtained registration is indeed a Teichm\"uller mapping. We also test the method to register cortical surfaces with more sulcal landmarks. In Figure \ref{fig:brainexample2},  we compute the Teichm\"uller extremal mapping between 2 brain surfaces with 6 corresponding sulcal landmarks labeled. The obtained registration and the norm of its associated BC is shown in (B) and (C) respectively. The landmarks are exactly matched. Again, the norm of the BC is a constant, showing that the obtained registration is a Teichm\"uller mapping which minimizes the conformality distortion.

\subsection{Constrained texture mapping}
\begin{figure*}[t]
\centering
\includegraphics[height=1.75in]{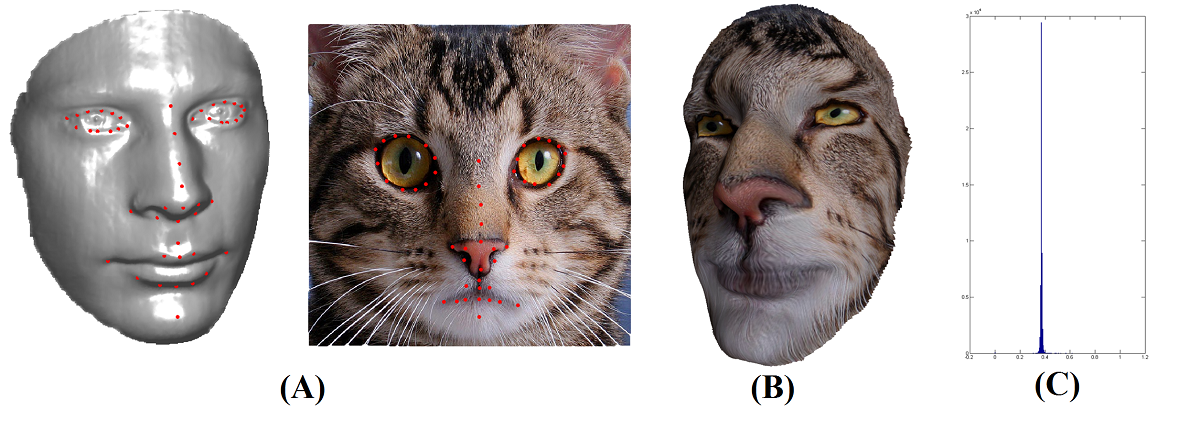}
\caption{(A) shows a human face and a texture image of a cat. Corresponding landmark points are labeled on the surface and the texture image. We compute the Teichm\"uller extremal mapping that matches the landmark points. The Teichm\"uller mapping is used as constrained texture mapping to project the texture image onto the surface, as shown in (B). (C) shows the histogram of the norm of BC. \label{fig:texture1}}
\end{figure*}

Texture mapping is one of the major photorealistic techniques in computer graphics to generate realistic and visually rich 3D surfaces \cite{Texturemappingintro1,Texturemappingintro2}. It is usually done by putting each surface mesh in correspondence with a 2D image \cite{Zhang,Hakertexture,Levy1,Levy2,SDMa,Bennis,Balmelli}. Such a correspondence between the surface mesh and the image is called the texture mapping. Constrained texture mappings are popularly used, in which the texture mappings are guided by landmark features labeled interactively by users. Ideally, the texture mapping should have minimum distortion, while matching the landmark points exactly. We apply our algorithms to compute the landmark matching Teichm\"uller extremal mapping between the surface mesh and the image, and use it as the texture mapping. In Figure \ref{fig:texture1}, we map a cat image onto the human face surface. (A) shows the corresponding landmark points labeled manually on the human face and the texture image. The landmark matching Teichm\"uller extremal mapping is computed, and is used as texture mapping to project the cat image onto the human face surface. The textured surface is as shown in (B). (C) shows the norm of the associated BC of the texture mapping. The norm is approximately a constant. It means the texture mapping computed is indeed a Teichm\"uller extremal mapping, which minimizes the conformality distortion.

\begin{figure*}[t]
\centering
\includegraphics[height=1.75in]{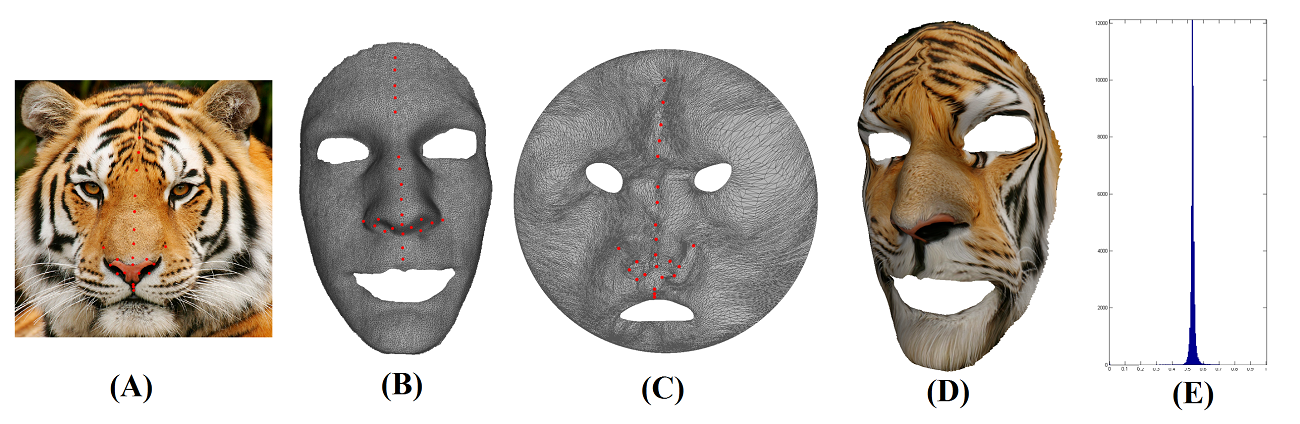}
\caption{(A) a texture image of a tiger. (B) shows a multiply-connected human face. Corresponding landmark points are labeled on the surface and the texture image. In (C), the surface mesh is mapped to a multiply-connected domain in 2D by a Teichm\"uller extremal mapping matching the landmark points exactly. The Teichm\"uller mapping is used as constrained texture mapping to project the texture image onto the surface, as shown in (D). (E) shows the histogram of the norm of BC. \label{fig:texture2}}
\end{figure*}

In Figure \ref{fig:texture2}, we further test our algorithm on a multiply-connected human face. (A) and (B) shows the texture(tiger) image and a multiply-connected human face. Corresponding landmark points are labeled manually on the texture image and the surface mesh. In (C), the surface mesh is mapped to a multiply-connected domain in 2D by a Teichm\"uller extremal mapping matching the landmark points exactly. (D) shows the textured surface. (E) shows the norm of the associated BC of the texture mapping. The norm is approximately a constant, which means the texture mapping computed is indeed a Teichm\"uller extremal mapping.

\subsection{Human face registration}
\begin{figure*}[t]
\centering
\includegraphics[height=1.65in]{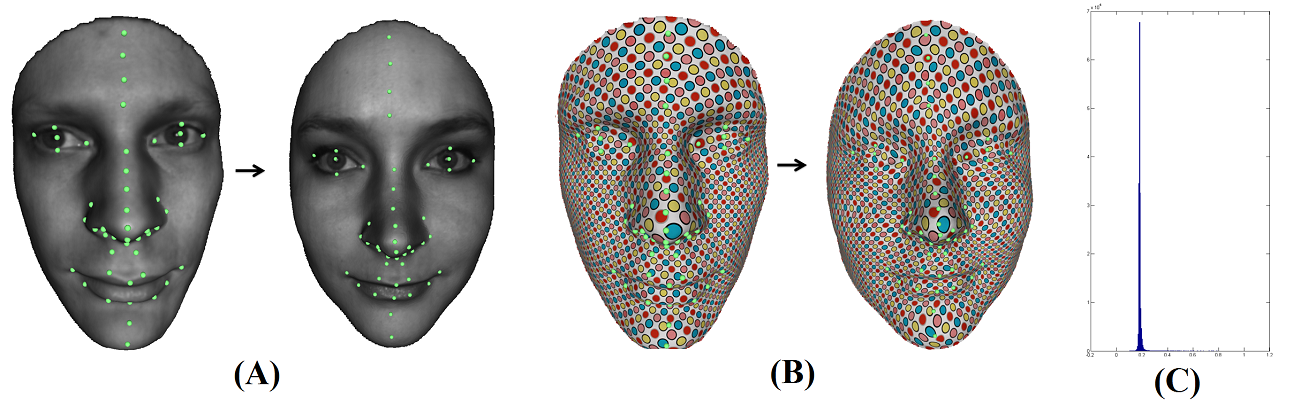}
\caption{Teichm\"uller extremal mapping of the simply-connected domain with landmark point constraints (A) shows the two faces with landmark point constraints. (B) shows the Teichm\"uller extremal mapping of the two faces. The resultant mapping is illustrated by texture mapping. (C) shows the histogram of the norm of BC. \label{fig:face1}}
\end{figure*}

In face recognition, finding accurate spatial correspondences between human faces is an a crucial process to compare and recognize faces effectively. Corresponding features can be extracted on human face based on curvatures, such as high curvature points near nose tips and lips. Accurate face registration can then be obtained by computing a mapping that matches the corresponding features. Landmark matching Teichm\"uller extremal mapping, which minimizes the geometric distortion, can then be used. In Figure \ref{fig:face1}, we apply our algorithm to compute the registration between a male and female human faces. The human faces are both simply-connected open surfaces. Corresponding feature points are labeled on both faces. The obtained Teichm\"uller mapping is obtained, which is visualized by texture mapping. The corresponding features are exactly matched. (C) shows the histogram of the norm of the BC, which is almost a constant. This demonstrates the obtained registration is a Teichm\"uller mapping.

\begin{figure*}[t]
\centering
\includegraphics[height=1.65in]{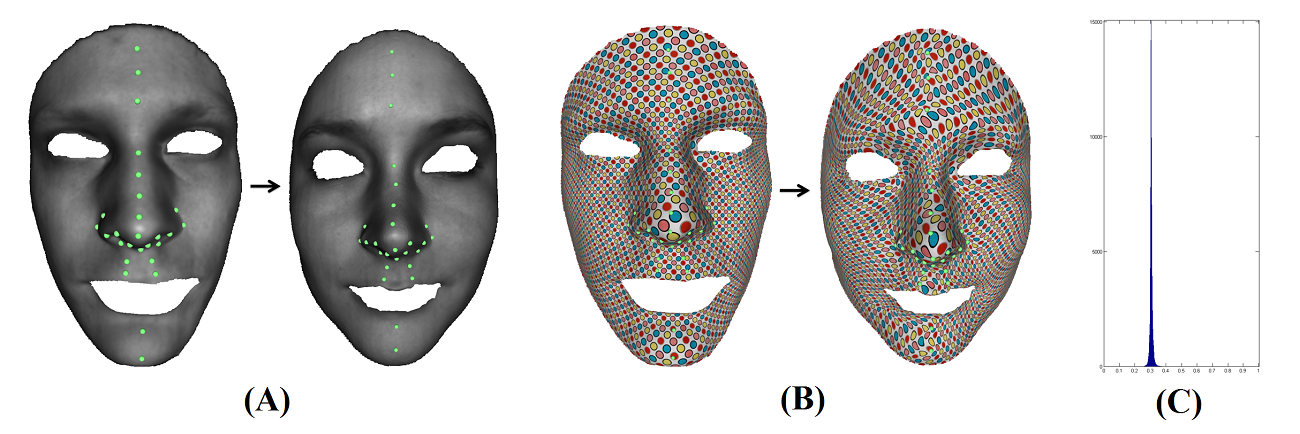}
\caption{Teichm\"uller extremal mapping of the multiply-connected domain with landmark point constraints (A) shows the two faces with landmark point constraints. (B) shows the Teichm\"uller extremal mapping of the two faces. The resultant mapping is illustrated by texture mapping. (C) shows the histogram of the norm of BC. \label{fig:face2}}
\end{figure*}

Our algorithm can also be applied to obtain registration between multiply-connected human faces. Figure \ref{fig:face2} shows two multiply-connected human faces. Corresponding feature landmarks are labeled. Teichm\"uller extremal mapping matching the features exactly is computed, as shown in (B). It is again visualized by texture mapping. (C) shows the histogram of the norm of the BC. Again, it is almost a constant, which demonstrates that obtained registration is a Teichm\"uller mapping minimizing the conformality distortion.

\section{Conclusion}
We address the problem of computing Teichm\"uller extremal mapping between surfaces, which minimizes the maximal conformality distortion. The proposed algorithm can be applied to obtain a landmark matching registration between surface meshes. Given a set of corresponding landmark points or curves defined on both surfaces, a unique landmark matching quasi-conformal registration can be obtained, which minimizes the conformality distortion. In this paper, we propose an efficient iterative algorithm, called the Quasi-conformal (QC) iterations, to compute the Teichm\"uller extremal mapping. The key idea is to represent the set of diffeomorphisms by Beltrami coefficients (BCs). We then look for an optimal BC associated to the desired Teichm\"uller
mapping. The associated diffeomorphism can be efficiently reconstructed from the optimal BC using the Linear Beltrami Solver(LBS). Using our proposed method, the Teichm\"uller mapping can be accurately and efficiently computed within 10 seconds. The obtained registration is guaranteed to be bijective. Besides, Teichm\"uller mapping with soft landmark constraints can also be computed using our proposed algorithm. It becomes useful when landmark features cannot be accurately located, and hence it is better to compute registration with landmarks approximately (but not exactly) matched. We applied the proposed algorithm to real applications, such as brain landmark matching registration, constrained texture mapping and human face registration. Experimental results shows that our method is effective in computing a non-overlap landmark matching registration with least amount of conformality distortion.

\end{document}